\documentclass[useAMS,usenatbib]{mn2e}
\usepackage[pdftex]{graphicx}

\title[Giant AGN clouds]{The Galaxy Zoo survey for giant AGN-ionized clouds: past and present black-hole accretion events}
\author[W.C. Keel et al.]
{William C. Keel$^{1,2,3}$\thanks{E-mail: wkeel@ua.edu}, S. Drew Chojnowski$^{2,3,4}$,  Vardha N. Bennert$^5,6$,
\newauthor  Kevin Schawinski$^{7,8,9}$,
Chris J. Lintott$^{10,11}$, Stuart Lynn$^{2,10}$, Anna Pancoast$^5$, 
\newauthor Chelsea Harris$^5$, A.M. Nierenberg$^5$, Alessandro Sonnenfeld$^5$, 
\& Richard Proctor$^{12}$\\
$^{1}$Department of Physics and Astronomy, University of Alabama,
Box 870324, Tuscaloosa, AL 35487, USA\\
$^{2}$Visiting Astronomer, Kitt Peak National Observatory,
operated by AURA, Inc.\ under contract to the US National\\ Science
Foundation.\\
$^{3}$ SARA Observatory, Physics and Space Sciences Department, Florida Institute of Technology, Melbourne, FL 32901, USA\\
$^{4}$Texas Christian University, Forth Worth, TX 76129 USA\\ 
$^{5}$ Department of Physics, University of California, Santa Barbara, CA 93106 USA\\
$^{6}$ Physics Department, California Polytechnic State University, San Luis Obispo, CA 93407, USA\\
$^{7}$Department of Physics, Yale University, New Haven, CT 06511 USA\\
$^{8}$Yale Center for Astronomy and Astrophysics, Yale University, P.O.Box 208121, New Haven, CT 06520, USA\\
$^{9}$Einstein Fellow\\
$^{10}$Astrophysics, Oxford University, Denys Wilkinson Building, Keble Road, Oxford OX1 3RH
\\
$^{11}$Adler Planetarium, 1300 S. Lakeshore Drive, Chicago, IL 60605\\
$^{12}$Waveney Consulting, Wimborne, Dorset BH21 3QY 
}

\begin{document}

\date{}

\pagerange{\pageref{firstpage}--\pageref{lastpage}} \pubyear{2011}

\maketitle

\label{firstpage}

\begin{abstract}
Some active galactic nuclei (AGN) are surrounded by extended
emission-line regions (EELRs), which trace both the illumination pattern of escaping
radiation and its history over the light-travel time from the AGN to the gas. From a new set of such
EELRs, we present evidence that 
the AGN in many Seyfert galaxies undergo luminous episodes 0.2--2$\times 10^5$ years in duration.
Motivated by the discovery of the spectacular
nebula known as Hanny's Voorwerp, ionized by a powerful AGN which has apparently faded dramatically
within $\approx 10^5$ years, Galaxy Zoo volunteers have carried out both targeted and serendipitous searches for 
similar emission-line clouds around low-redshift galaxies. We present the resulting list of candidates
and describe spectroscopy identifying 19 galaxies with AGN-ionized regions at projected radii $r_{proj} > 10$ kpc.
This search recovered known EELRs (such as Mkn 78, Mkn 266, and NGC 5252) and identified
additional previously unknown cases, one with detected emission to $r=37$ kpc. One new Sy 2 was
identified.  At least 14/19 are in interacting or merging systems, suggesting that tidal tails are
a prime source of distant gas out of the galaxy plane to be ionized by an AGN. We see a mix of one- and two-sided structures, 
with observed cone angles from 23--112$^\circ$. We consider the energy balance in the
ionized clouds, with lower and upper bounds on ionizing luminosity from recombination and ionization-parameter arguments,
and estimate the luminosity of the core from the far-infrared data. The implied ratio of ionizing radiation seen by the clouds
to that emitted by the nucleus, on the assumption of a nonvariable nuclear source, ranges from 0.02 to $>12$; 7/19 exceed unity. Small values
fit well with a heavily obscured AGN in which only a small fraction of the ionizing output escapes to be traced by surrounding
gas. However, large values may require that the AGN has faded over tens of thousands of years, giving us several examples
of systems in which such dramatic long-period variation has occurred; this is the only current technique for addressing these
timescales in AGN history. The relative numbers of faded and non-faded objects we infer, and the projected extents of the
ionized regions, give our estimate (0.2--2$\times 10^5$ years ) for the length of individual bright phases. \end{abstract}


\section{Introduction}
The compact sizes of the central engines of active galactic nuclei (AGN) have long driven
study of their distant surroundings for clues to their geometry and interaction with
the surrounding galaxy. Observations of gas seen many kpc from the AGN itself
have proven fruitful in offering views of the core from different angles, and implicitly at
different times.

Narrowband images revealed extended emission-line regions (EELRs) around some luminous
AGN, particularly radio-loud QSOs as well as radio galaxies, as reviewed by \cite{Stockton2006}.
Similar structures in lower-luminosity Seyfert galaxies often appear as single or double triangles in projection
(\citealt{Unger1987}, \citealt{Tadhunter1989}), generally interpreted as ionization cones.  When 
small-scale radio jets are present, they lie within the ionization cones. 
However, in many cases,
the gas must be ionized by radiation from the nucleus rather than direct
interaction with a jet or outflow, as seen from narrow linewidths and (particularly diagnostic)
modest
electron temperatures, both of which would be much larger in the presence
of shocks fast enough to match the observed ionization levels. This is particularly true for
very large EELRs, where interaction with the radio jet or an origin in outflows alone
become less and less likely. In fact, the best-defined ionization cones are
seen in radio-quiet objects \citep{Wilson1996}.

This is one line of evidence linking large-scale structures to
the small-scale obscuring regions (``tori") implied by other arguments for a unification scheme
\citep{Ski93}, in
which Seyferts of types 1 and 2 are part of a single parent population, appearing different based
on how our line of sight passes this torus. The emission-line structures can be large and well-resolved,
offering a way to measure the opening angle over which ionizing radiation escapes. Some previous
studies have also noted that these emission-line clouds provide a view to the immediate past of the
AGN, via light-travel time to the cloud and then toward us \citep{Dadina}.

Using extended emission-line clouds as probes of AGN history came of age with
the discovery of Hanny's Voorwerp, a high-ionization region extending 45 kpc
in projection from the low-ionization nuclear emission-line region (LINER) galaxy 
IC 2497 at $z=0.05$ \citep{Lintott2009}. Linewidths and electron
temperature indicate that the gas is photoionized rather than shock-excited,
while a combination of ionization-parameter and recombination arguments 
bound the required nuclear ionizing luminosity to be $1-4 \times 10^{45}$ 
erg s$^{-1}$. However, X-ray spectroscopy shows the nucleus of IC 2497
to be only modestly absorbed, with ionizing luminosity only $\approx 10^{42} $
erg s$^{-1}$ \citep{Schawinski2010a}. It is difficult to avoid the conclusion that
the nucleus of IC 2497 was in fact a QSO (the nearest known luminous QSO)
until roughly $10^5$ years before our current view, and has faded 
dramatically in the interim; radio and HST observations offer hints that 
some of its energy output may have switched to kinetic forms over
this timespan (\citealt{Josza2009}, \citealt{Rampadarath}, \citealt{Schawinski2010a}, Keel
et al. in preparation).
The unlikeliness of the nearest QSO showing highly unusual behaviour 
suggests that such variations may be common among AGN, prompting us to 
re-examine the incidence and properties of extended ionized clouds around 
nearby AGN. Such an examination should not be confined to catalogued AGN, since
the most interesting objects - those which have faded dramatically - may no
longer appear as spectroscopically classified AGN.

Hanny's Voorwerp was first noted by Dutch teacher Hanny van Arkel in the course of
the Galaxy Zoo project \citep{Lintott2008}, on the basis of its unusual structure and colour.
In view of the interest of similar ionized clouds for study of both the history and
obscuration of AGN, participants in the Galaxy Zoo project have carried out a wide search 
for such clouds using data from the Sloan Digital Sky Survey (SDSS). They 
examined both known AGN hosts and galaxies not known to have AGN, using the
distinctive colour of highly-ionized regions across the SDSS $gri$ filters as a first
selection criterion. We present the results of further analysis of the SDSS images, 
narrow-band imaging, and spectroscopy, yielding a list of 19 galaxies with AGN-photoionized clouds
detected to beyond 10 kpc from the nuclei (many of which are newly identified). We 
consider constraints on changes in
ionizing luminosity for these, and identify several as the most likely candidates for
the kind of long-term fading seen in IC 2497 and Hanny's Voorwerp. 

\section{Searches for emission-line clouds}

The Galaxy Zoo search for giant AGN-ionized clouds combined both targeted
and serendipitous approaches, to combine a complete examination of known AGN hosts with
the possibility of finding ionized clouds around AGN which are yet unknown or in fact optically
unseen. In the targeted search, we formed a sample of
potential AGN at $z<0.1$. This combined all galaxies whose
SDSS pipeline emission-line ratios put them in either the AGN or composite regions of the Baldwin-Phillips-Terlevich (BPT)
diagram (\citealt{BPT},
as revised by \citealt{Kewley2001} and \citealt{Kauffmann2003}) using [O III]/H$\beta$ and [N II]/H$\alpha$, and all additional objects listed in
the Veron-Cetty \& Veron catalog \citep{VCV13} at $z<0.1$ falling within the SDSS
data release 7 (DR7) area. This addition accounted for AGN with no SDSS
nuclear spectrum, either because they are relatively bright or, more often, because fibre
collisions or sampling rules prevented their selection for spectra, and
type 1 AGN where the pipeline spectroscopic classification is less reliable than for narrow-line objects. The merged
AGN sample, designed to err on the side of inclusion in borderline cases, included 18,116 objects. With a web interface designed by RP, 
199 participants examined all of these within a 6-week period in 2009,
marking each as certain, possible, or lacking an extended emission region.
These emission regions have distinctive signatures in both morphology and colour from the SDSS data.
They do not follow the usual spiral or annular distributions of star
formation in disc galaxies. Such regions show unusual colours in the SDSS
composite images, which map $gri$ bands to blue, green, and red (Lupton et al. 2004). Hence
strong [O III] at low redshift is rendered as a pure blue, as in the discovery of Hanny's Voorwerp. A combination of strong [O III] and significant H$\alpha$+[N II] appears purple; beyond about $z=0.1$, [O III] falls in the gap between $g$ and $r$ filters, so our search technique loses utility until [O III] is well within the $r$ band, when the galaxies have much smaller angular sizes. This subproject was known as the ``voorwerpje hunt", using the Dutch diminutive form of Voorwerp.

Each galaxy was examined by at least ten participants; 199 Zoo volunteers participated in this program,
seven of whom examined the entire sample. The final average number of votes was 11.2 per object. After this screening process, a straightforward ranking was by relative numbers of ``yes" (weight=1), ``maybe" (weight=0.5), and ``no" (weighted zero) votes.

The most interesting results of such a search would be galaxies with
prominent AGN-ionized clouds in which we don't see the AGN, either because
of strong obscuration or dramatic variability during the light-travel time from the nucleus to the clouds. These would not be found by targeting known AGN, and neither would clouds around AGN which do not have catalogued spectral information. Accordingly, we also posted a request on the Galaxy Zoo discussion forum,
with examples of confirmed AGN clouds and various kinds of similar-appearing image artifacts. Participants were invited to post possible cases from among the galaxies they saw in the ordinary course of the Galaxy Zoo classification programs \citep{Lintott2008}, and some active users reposted examples from
other discussion threads. The resulting followups\footnote{in http://www.galaxyzooforum.org/index.php?topic=275014.0} provided an additional sample for investigation; WCK also checked all the threads with image
discussion for more such objects, early examples of which instigated this search on the first place. To reduce the number of false positives caused by extended star-forming regions or starburst winds, objects were removed
from consideration if an SDSS spectrum shows emission lines characteristic of a starburst. Remaining candidates were examined first on the SDSS composite images for appropriate colour and geometry, and the most promising ones were carried forward for further analysis.

Both targeted and serendipitous lists overlap for many objects with bright
emission-line structures, and recover such well-studied cases from the literature as Mkn 266, NGC 5252 and Mkn 78; we observed these so as
to have a consistent set of spectra for comparison. The entire list of candidates
is given in Table \ref{tbl-candidates}. In the Search column, S or T denotes whether the object was found in the serendipitous survey, the targeted
survey of known AGN, or both. The type of nuclear optical spectrum is listed as Sy 1/1.5/1.9/2, LINER, SB for starburst, or nonAGN for an ordinary stellar
population. The final column indicates which Galaxy Zoo participant (by user name) first posted objects in the serendipitous survey.

\subsection{SDSS image analysis and new images}

\begin{figure} 
\includegraphics[width=67.mm,angle=270]{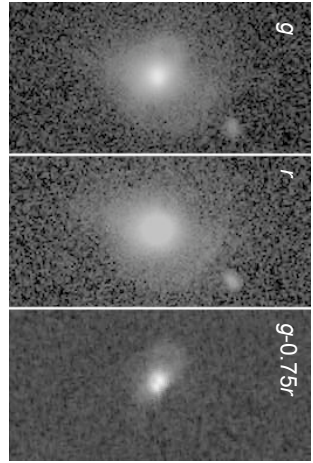} 
\caption{Linear combination of SDSS images to isolate candidate [O III] emission regions, shown
with SDSS J143029.88+133912.0 (the Teacup AGN). The region shown spans $64 \times 128$ SDSS pixels,
or $25.3 \times 50.6$", with north at the top. Each image is displayed with a logarithmic intensity 
mapping, with a small zero point
offset to reduce the effects of noise around zero. At the redshift $z=0.085$ of this galaxy,
H$\alpha$ and [N II] emission fall redward of the $r$ filter band (response 0.2\% of peak) 
so the $r$ image is used as a continuum estimate. }
 \label{fig-sdssteacup}
\end{figure}

For both subsamples, further winnowing had the same steps. Most importantly,
we reanalyzed the SDSS images, to verify that the features do not have continuum counterparts, and eliminate artifacts caused by imperfect registration of the images when forming the colour composites. This effect
is of particular concern for Seyfert 1 nuclei, where the PSF of the bright nucleus can produce a decentered colour signature if one of the constituent images is slightly misregistered; Sy 1 galaxy image are more vulnerable to this artifact than normal galactic nuclei. Since many candidates (including some with spectroscopic confirmation) have ``purple haze" on the SDSS images, which could either be genuinely extended and somewhat amorphous [O III] and H$\alpha$ or an artifact, this was a helpful step. We adopted a tomographic
approach, taking one of the SDSS bands free of strong emission lines ($r$ or $i$, depending on redshift) as an estimate of the structure of starlight in the galaxy. This was scaled to match the largest part of the $g$ structure, iteratively when necessary. This is illustrated in Fig. \ref{fig-sdssteacup}, isolating the
emission-line loop in SDSS 1430+13 (nicknamed the Teacup AGN because of this structure). Chojnowski and Keel inspected the best subtraction among various scalings
(often a compromise, due to colour gradients within the galaxy) to assess 
the reality of extended emission-line features not associated with clear spiral arms or stellar rings. These results let us rank the candidate
lists from both targeted and serendipitous searches in order of significance of
the emission-line structures based on the SDSS images themselves. We used these results to limit the number of candidates from the targeted
search to the top 50; below this there were no convincing candidates based on more detailed analysis of the SDSS images.

\begin{figure} 
\includegraphics[width=67.mm,angle=90]{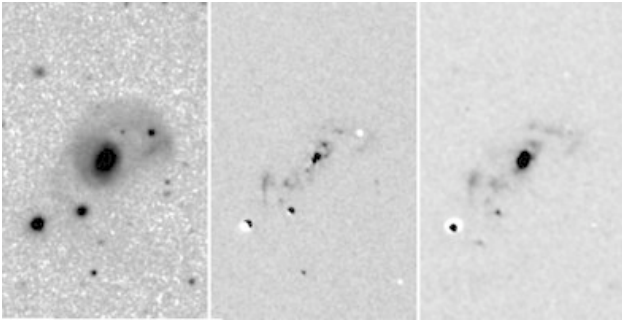} 
\caption{The extended clouds in UGC 7342. Left, the starlight continuum in a band at 6450 \AA\   from the SARA 1m telescope. Center,
an estimated [O III] image from the SDSS data as in Fig. \ref{fig-sdssteacup}. Right, continuum-subtracted H$\alpha$ image
from the SARA 1m. North is at the top and east to the left; the field shown spans $97 \times 150$ arcseconds.}
 \label{fig-sarapix1}
\end{figure}

\begin{figure} 
\includegraphics[width=67.mm,angle=270]{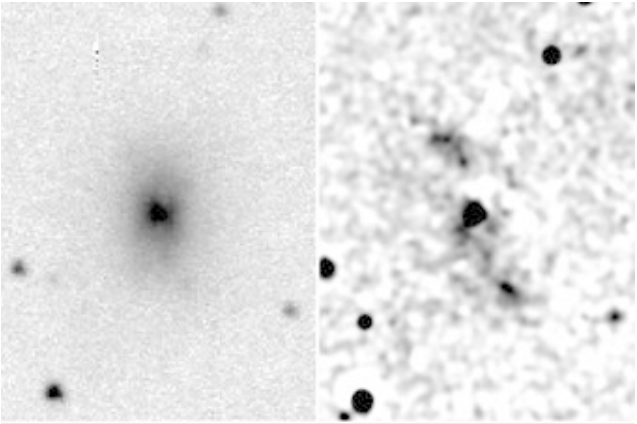} 
\caption{The extended clouds in SDSS 2201+11. Left, a $g$ image from the SARA-S 0.6m telescope, showing the dusty disk.
Right, continuum-subtracted [O III] image from the SARA-N 1m telescope, smoothed by a Gaussian of 2.0" FWHM.
North is at the top and east to the left; the field shown spans $64 \times 84$ arcseconds.}
 \label{fig-sarapix2}
\end{figure}

Where appropriate filters were available for [O III] or H$\alpha$
at a galaxy's redshift, some candidates were imaged at the remote SARA
1m (Kitt Peak) and 0.6m (Cerro Tololo) telescopes. For [O III], we
used a filter centered at 5100 \AA\  with half-transmission width 100\AA\ ,
useable for the redshift range $z= 0.009 - 0.025$. At H$\alpha$, both telescopes have stepped sets of filters 75 \AA\  apart with FWHM $\sim 75$ \AA\ . Continuum
was taken from $V,R$ or $g,r$, appropriately scaled for subtraction to show net emission-line structures. These data are particularly helpful in tracing the emission-line structures of UGC 7342 (Fig. \ref{fig-sarapix1}) and SDSS 2201+11 (Fig. \ref{fig-sarapix2}).

\subsection{Spectroscopy}

To confirm that regions are in fact ionized by AGN, and derive diagnostic emission-line properties, we carried out long-slit spectroscopy for the highest-priority candidates. Observations used the GoldCam
spectrograph at the 2.1m telescope of Kitt Peak National Observatory and the Kast double spectrograph at the
3m Shane telescope of Lick Observatory.
Table \ref{tbl-observations} compares the setups used for each session.
The slit width was set at 2" for all these observations, and the spectrographs
were rotated to sample the most extended known structures of each galaxy.
Scheduling allowed us to reduce the Kitt Peak data before the first Lick observing run, so that the 3m spectra could be concentrated on the most
interesting galaxies. Total exposures ranged from 30 minutes, for initial
reconnaissance to see whether an object might host AGN clouds, to
2 hours for weaker lines in confirmed targets. Either night-sky line or interspersed lamp observations
were used to track flexure, as needed.  Reduction used the {\it longslit} package in IRAF\footnote{IRAF is distributed by the National Optical Astronomy Observatory, 
which is operated by the Association of Universities for Research in Astronomy (AURA) under cooperative agreement with the National Science Foundation.}
 \citep{Tody}, and included
rebinning to a linear wavelength scale, sky subtraction, and flux calibration.
Spectrophotometric standard stars were observed to set the flux scale; in a few cases where passing
clouds were an issue, the spectra were scaled so that the nucleus within a $2 \times 3$" region matched the
flux of the SDSS spectra.

\begin{figure*} 
\includegraphics[width=140.mm,angle=270]{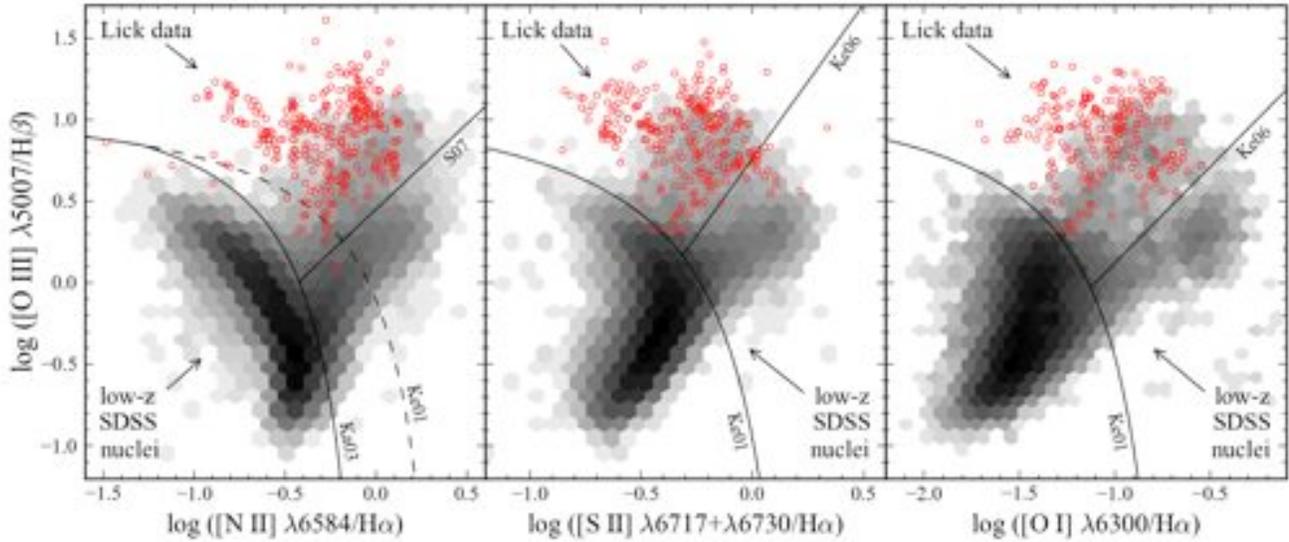} 
\caption{Summary Baldwin-Phillips-Terlevich (BPT) diagrams for the Lick spectra, where we measured the requisite red emission lines.
Circles indicate points along the slit for extended clouds classified as AGN-ionized and the host nuclei. Gray-scale
background shows the density of points from all low-redshift SDSS galactic nuclei, as in Schawinski et al. (2010b). The dividing lines
between regions photoionized by AGN and by hot stars are shown as given by Kewley et al. (2001; Ke01) and by Kauffmann et al. 
(2003; Ka03).}
 \label{fig-bptkey}
\end{figure*}

Our identification of these extended regions as being photoionized by AGN rests on three results - location in the strong-line
BPT diagram, strength of the high-ionization species He II and [Ne V], and electron temperature consistent with photoionization
but not with shock ionization.
We classify emission regions based on the ``BPT" line-ratio diagrams pioneered for galactic nuclei by \cite{BPT}
and refined by \cite{VO87}, with caution based on the possibility that some of the external gas could have much lower metal abundances than found in galactic nuclei (as seen in Hanny's Voorwerp; \citealt{Lintott2009}). Abundance effects in gas photoionized by AGN, as manifested in the BPT diagrams, have been considered in calculations
by \cite{Bennert2006a}. The largest effect is higher equilibrium temperature at lower O abundance, since it is an important coolant, which drives
stronger forbidden lines and higher ionization levels until very low levels (0.1 solar) are reached. In any case,
the abundance changes are not large enough to move these clouds across the empirical AGN/starburst ionization boundary. 
Furthermore, in the galaxies where we have data covering the red emission lines, the clouds' locations in the
(essentially abundance-independent) auxiliary BPT diagram of [O III]/H$\beta$ versus [O I]/H$\alpha$ also indicated
photoionization by an AGN continuum. The various BPT diagrams are compared for 
Points along the slit in each of the clouds we classify as AGN-ionized  in Fig.
\ref{fig-bptkey}. This classification
is examined more closely in the context of its radial behavior in the next section. 

Independent of these
line ratios, strong He II $\lambda 4686$ or [Ne V] $\lambda \lambda 3346, 3426$ indicate photoionization by a harder continuum than provided by young
stars, and resolved emission from these species is immediately diagnostic of AGN photoionization in this context. For some objects, we do not have
red data; in these, we classify the cloud as AGN-ionized based on the presence of the high-ionization lines or continuity of line ratios with
the nucleus. Line ratios in the extreme blue may be affected in subtle ways by atmospheric dispersion \citep{AVF82}; the scheduling of our
observations forced us to observe most targets at hour angles which did not allow us to put the slit simultaneously along the structures of interest and 
close to the parallactic angle. The
extended regions we observe are generally wider than the slit; to first order line intensities are not affected by atmospheric refraction,
since we calibrate with standard stars at low airmass. Some of the Lick blue spectra have atmospheric dispersion contributing 
as much as 3" of offset along the slit from red to blue ends of the spectrum, important only for the nuclei and corrected in extracting their spectra.

The BPT diagrams are designed to separate common sources of
photoionization in galaxies; temperature and kinematic data are also important to understand whether shocks pay a significant role.
In a few cases, the [O III] $\lambda 4363$ line was measured in the extended clouds with sufficient
signal-to-noise ratio for a measurement of the electron temperature via its ratio to the strong
$\lambda \lambda 4959, 5007$ lines. Using the IRAF
task {\it temden}, which implements the algorithm of \cite{ShawDufour}, and considering 
$n_e < 100$ cm$^{-3}$, we find $T_e$ values of $18,600 \pm 1000$ in the
SDSS 2201+11,  $13,300 \pm 300$ for Mkn 266, and $15,400 \pm 500$ for the Teacup system.
These confirm that the gas is photoionized rather than shocked; for comparison, temperatures in the [O III]
zone of supernova remnants (including some with lower ionization levels than in these clouds)
range from 20,000-69,000 K (e.g., \citealt{Fesen82}, \citealt{Wallerstein}, \citealt{Morse95}).
In addition, very high shock velocities $ \approx 400$ km s$^{-1}$ are needed to produce significant [Ne V] emission
\cite{Dopita96}. This is far in excess of the local velocity ranges we observe (section 6); even though we would not necessarily observe
material on both sides of a shock in the same ion, it is difficult to envision a situation with large-scale
shocks of this velocity without observable velocity widths or structures exceeding 100 k s$^{-1}$.

While not the main thrust of our survey, it is worth noting that we find a few instances of either double AGN in interacting systems, or
AGN in the fainter member of a close pair (Mkn 177, Was 49, possibly SDSS J111100.60-005334.9 and
SDSS J142522.28+141126.5). These may be worth deeper spectroscopy in the context of mapping AGN obscuration;
if a high-ionization component can be isolated in the gas of the 
other galaxy, its distribution could show where ionizing radiation escapes any circumnuclear absorbing structure.
This offers a distinct way of tracing the ionizing radiation even in the absence of extensive gaseous tidal features, in
an approach that has been discussed for Was 49ab by \cite{Moran}.

\begin{figure*} 
\includegraphics[width=180.mm,angle=0]{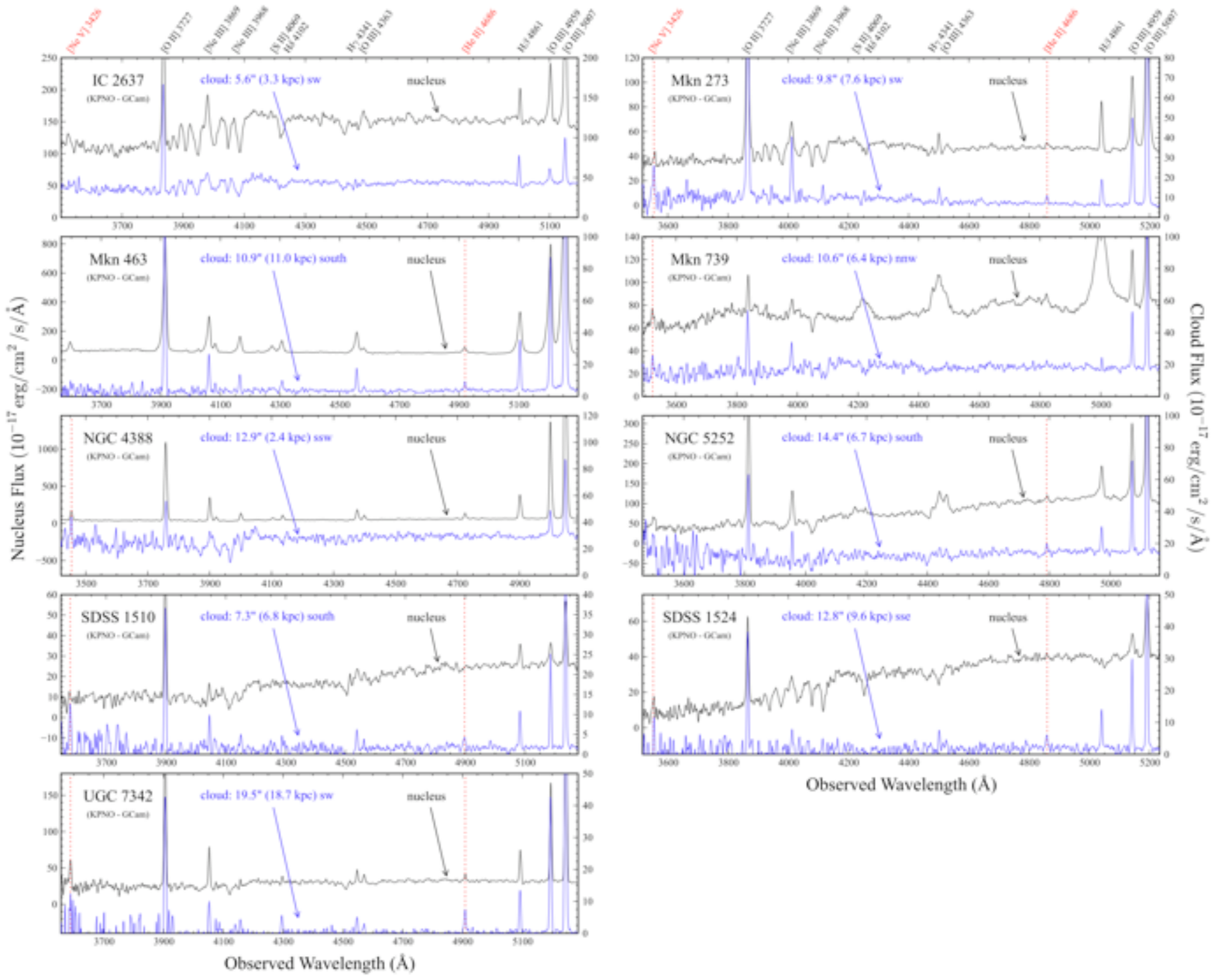} 
\caption{KPNO GoldCam  spectra sampling nuclei and associated AGN-ionized clouds. 
Spectra are scaled to show [O III] $\lambda 4959$, with nuclear spectra representing $2 \times 3.1$-arcsecond areas and cloud spectra representing 
$2 \times 6.2$-arcsecond areas.
He II and [Ne V] emission, especially important as indicators of a hard ionizing radiation field, are indicated by red dotted lines when clearly detected in clouds. Distances and directions of cloud relative to nuclei are indicated as shown.}
 \label{fig-kpnospectra}
\end{figure*}

Table \ref{tbl-confirmations} lists the results of our spectroscopy.
Confirmed, resolved clouds ionized by the AGN are separated from other
results (unresolved AGN emission, extended star-forming regions denoted as H II, and so on). The
instruments used are denoted by GCam (Kitt Peak GoldCam) and Lick (Lick 3m with Kast spectrograph).
New redshifts and spectral classifications are marked with asterisks. We separate the AGN clouds of
most interest based on the detected extent of [O III] $\lambda 5007$; our spectra have a lower
detection threshold than our images for this, roughly $10^{-16}$ erg cm$^{-2}$ s$^{-1}$ arcsec$^{-2}$
for emission regions a few arcseconds in size. Spectra of the nuclei and representative cloud regions
are shown in Figs. \ref{fig-lickspectra} and \ref{fig-kpnospectra}. Table \ref{tbl-lineratios} lists emission-line
ratios and selected fluxes for the same regions plotted in these figures. Fluxes are given both for [O III] $\lambda 5007$ and
H$\alpha$, since these were usually measure with different gratings and detectors. For some of the nuclei, 
correction of the H$\beta$ flux for underlying absorption in the stellar population was significant; we have applied
an approximate correction based on typical values for synthetic stellar populations from \cite{Keel83}.

\begin{figure*} 
\includegraphics[width=180.mm,angle=0]{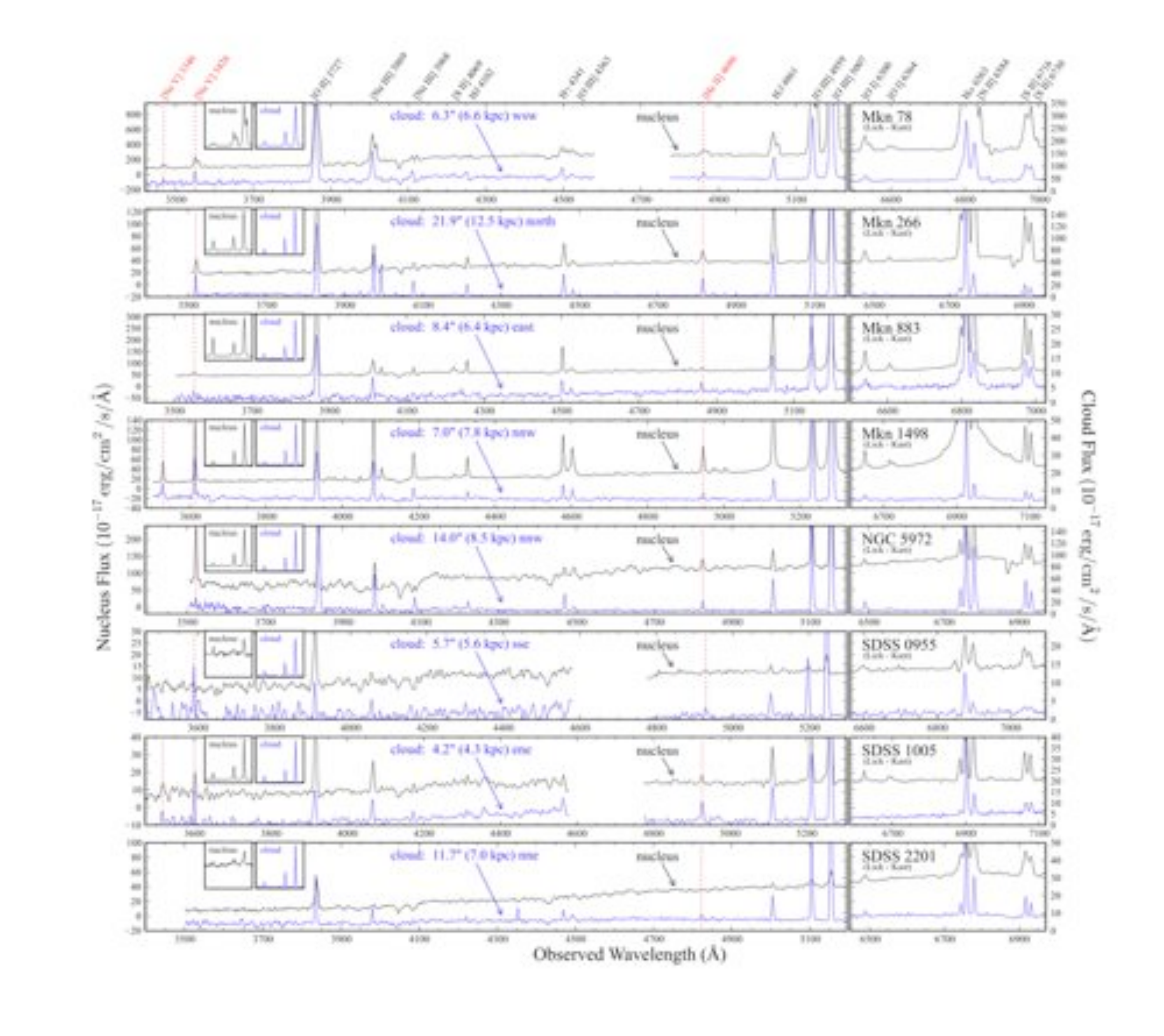} 
\caption{Lick spectra for nuclei and associated AGN-ionized clouds.
Small insets at left show the H$\beta$+[O III] region for nuclei and clouds, scaled to the peak of $\lambda 5007$ emission.
Panels on the right show the [O I] - [S II] region at the same flux scale as the blue spectra.
He II and [Ne V] emission, especially important as indicators of a hard ionizing radiation
field, are indicated by red dotted lines when
clearly detected in clouds. Nuclear spectra represent $2  \times 3.1$-arcsecond areas, and cloud spectra are summed over
$2 \times 6.2$-arcsecond areas. Distances and directions of cloud relative to nuclei are indicated as shown. Three spectra have
gaps in the blue region, since they were taken with the dichroic splitting red and blue optical
trains near 4600 \AA\  .}
 \label{fig-lickspectra}
\end{figure*} 

\begin{figure*} 
\includegraphics[width=180.mm,angle=0]{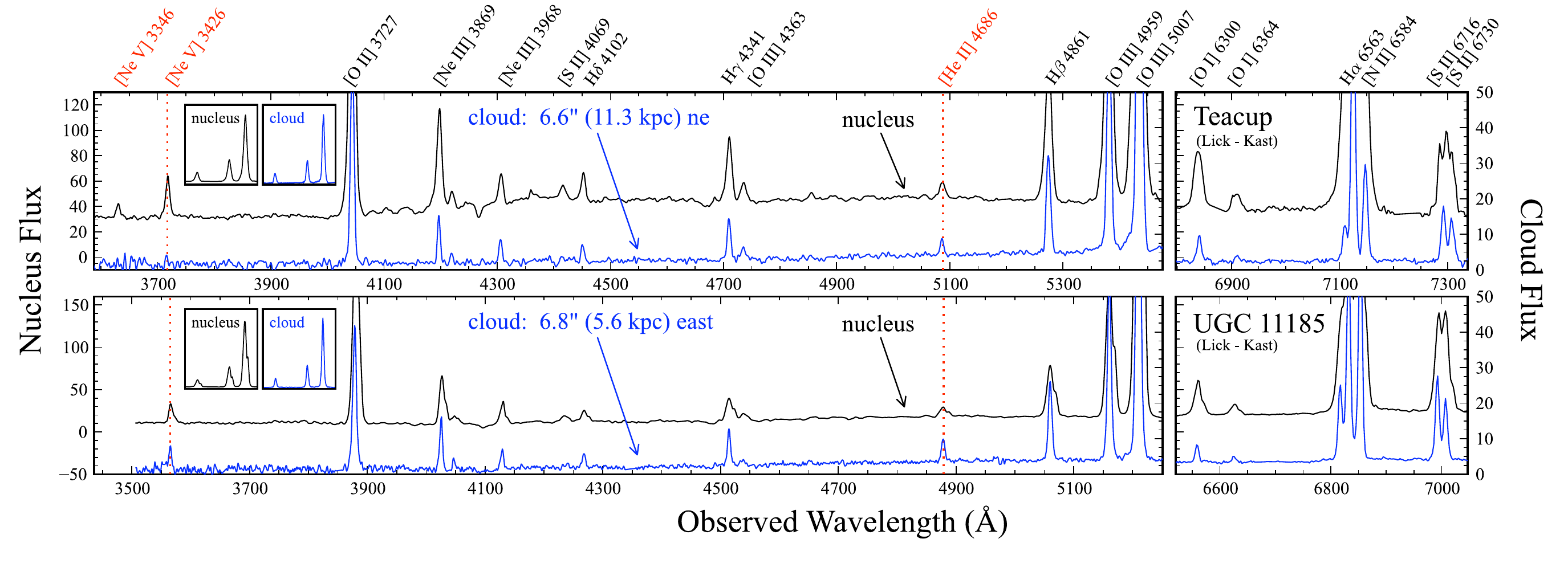} 
\contcaption{} 
\label{fig-lickspectra2} 
\end{figure*}

The upper part in Table  \ref{tbl-confirmations}, with AGN-ionized gas detected more
than 10 kpc from the nucleus, forms the sample for our subsequent analysis. As a sign of completeness, of these 19, 
14 were found in both targeted and serendipitous searches. SDSS J095559.88+395446.9 was newly identified as a 
type 2 Seyfert in our spectrum, after
having been found in the ``blind" search of galaxies independent of prior classification as
an AGN host (so it was not included in the targeted sample). Of the remainder, Mkn 78 and Mkn 463 were selected in the
targeted AGN sample, while Mkn 1498 and UGC 11185 were recognized only in the serendipitous survey.
It may be relevant that both Mkn 78 and Mkn 463 have ionized regions with relatively small projected extent,
easily lost against the galaxy starlight (which in Mkn 463 is morphologically complex).

The [S II] $\lambda 6717, 6731$ lines are particularly important, tracing electron densities and thereby
providing one estimate of the intensity of the impinging ionizing radiation. Since the densities
in these extended clouds are low, and the ratio is generally near its low-density limit, where
the mapping from line ratio to density is highly nonlinear, we have
examined the errors in measuring the line ratio closely. We generated multiple realizations of
pixel-to-pixel noise, and each was scaled to four representative fractions of the stronger line peak.
This was added to line pairs, modeled to match the line widths and pixel separation of the red
Lick data. Gaussian fitting of the lines with added noise gave a relation between the peak
signal-to-noise and error of the fitted ratio which we adopted; we use $\pm 2 \sigma$ error 
bounds to derive bounds on the density. Density values were calculated using the IRAF
task {\it temden}.

\section{Energy budget in extended clouds: obscuration versus variability}

Seeing the effects of radiation from an AGN on gas tens of kpc from the nucleus allows us the
possibility of tracing dramatic changes in core luminosity.
One straightforward way to approach this question is a simple energy balance. The spectra give
us upper and lower bounds on the required ionizing luminosity. To probe 
the most extreme conditions, we analyze galaxies in which we detect ionized
gas at projected distances $r > 10$ kpc. For all distances and luminosities,
we use the WMAP ``consensus" cosmological parameters, with H$_0 = 72$ 
km s$^{-1}$ Mpc$^{-1}$ \citep{Spergel}.

The lower bound comes from the highest recombination-line surface brightness
we observe; the central source must provide at least enough ionizing photons to sustain
this over periods longer than the recombination timescale (which may be as long as
$10^4$ years at these low densities). This is a lower limit, since the actual
emission-line surface brightness of some regions may be smeared out by seeing, and
we do not know that a given feature is optically thick at the Lyman limit. This
limit depends only very weakly on the slope of the ionizing continuum, since helium
will generally absorb most of the radiation shortward of its ionization edge leaving
only the 13.6-54.4 eV range to consider for hydrogen ionization. We base our
bounds on the highest implied luminosity among structures at various projected
radii in a given system, with no correction for projection effects. This makes our
limits conservative, since a given cloud will always lie farther from the nucleus than
our projected measurement. In essence, this argument is based on the surface brightness in
a recombination line; we use H$\beta$ since we have these data for the whole sample.
In a simple approximation, we take the surface brightness in the brightest portion of a
cloud, assuming this to be constant across the slit. We take the region sampled in this way to 
be circular in cross-section as seen from the nucleus, so its solid angle is derived from the
region subtended by the slit. We then see this region occupying a small angle $\alpha = 2 \arctan ({\rm slit ~half-width} /r) $ as seen
projected at angular distance $r$ 
from the AGN, the required ionizing luminosity is given from observed quantities as
$ L_{ion} = 1.3 \times 10^{64} z^2 F(H \beta)/\alpha^2 $  for $\alpha$ in degrees.
The derived values are listed in Table \ref{tbl-energy}, along with complementary quantities
related to the nuclear luminosity (as collected below). The derived ionizing luminosities are lower limits, since
there may be unresolved regions of higher surface brightness, and we do not know whether a given cloud is
optically thick in the Lyman continuum. Higher-resolution imaging in the emission lines could this increase these
values.

Upper limits to the incident ionizing flux come from a complementary analysis using the ionization parameter ($U$, the ratio of ionizing
photons to particles), since these emission-line features all have [S II] line ratios near the low-density limit.
Our density results from the $\lambda 6717 / \lambda 6731$ [S II] line ratio are given in Table
\ref{tbl-s2density}. Values are listed only for objects with useful measures far from the core.
In each case, we evaluated the density at a typical temperature of $10^4$ K, and at the
higher temperature $1.3 \times 10^4$ K found in Hanny's Voorwerp \citep{Lintott2009}
and in our data for Mkn 266 and SDSS 2201+11, where the higher temperature is set by thermal equilibrium for substantially subsolar oxygen abundance. We quote the extreme
range of density values between these two cases (allowing in the Teacup an upper bound
on the electron density as high as
240 cm$^{-3}$, and in some cases limits $<10$cm$^{-3}$), since the temperature-sensitive
[O III] $\lambda 4959+5007 / \lambda 4363$ 
line ratio is not well-enough measured in most of these objects to use individual $T_e$ values.
We derive $U$  from the [O II] $\lambda 3727$/[O III] $\lambda 5007$ ratio
using the power-law continuum models from \cite{KomossaSchulz}, and the analytic 
fits from \cite{Bennert2005} as interpolation tools. For fully ionized hydrogen at a distance $d$ from the
AGN, the photon flux in the ionizing continuum is $Q = 4 \pi d^2 n_e U/c$.
For objects with red spectra, giving densities from the [S II] lines, limits to the luminosity are given in 
Table \ref{tbl-s2density}.
It is reassuring that the upper limits to ionizing luminosity derived from $U$ and $n_e$ always fall above the lower 
limits from recombination balance.

The lower limits from recombination-balance are independent of assumptions about
the local density $n_e$, making it more robust than ionization-parameter arguments
when we have no independent tracer at these low densities. Fig. \ref{fig-BPTplots}
shows several of our objects in one of the ``BPT" diagrams, going beyond their initial
use to classify the gas as AGN-ionized to examine changes with projected distance from the nuclei.
Some of these, such as Mkn 1498 and the Teacup 1430+13, 
show a phenomenon remarked
earlier in, for example, NGC 5252 \citep{Dadina} - the ionization balance stays roughly
constant with radius, which is naturally explained if the characteristic density $n_e \propto r^{-2}$.
This might occur naturally for gas in the host galaxy; tidal streams of gas would not be likely to match the
extrapolated behavior of gas within the galaxy and indeed we see some cases (Mkn 266, NGC 5972, SDSS 2201+11) with
substantial radial changes in $U$. However, for Seyfert narrow-line regions, \cite{Bennert2006b} find a shallower
density gradient $n_e \propto R^{-1.1}$, which would imply $U \propto R^{-0.9}$ for gas which is optically thin (or has
a small covering fraction). These objects have heterogenous behavior;
In the ionization cone of NGC 7212, \cite{Cracco} find no radial trend of $n_e$.

Similar conclusions come from the more limited blue-line diagram also considered by \cite{BPT},
which we can apply to the objects for which we have only blue-light spectra from KPNO. Some
of objects in this diagram as well as in Fig. \ref{fig-BPTplots}  show systematic changes in ionization level 
with radius, manifested as offsets from
upper left (higher ionization) to lower right (lower ionization). We show this behavior in
Fig. \ref{fig-bluebptplot}.

\begin{figure*} 
\includegraphics[width=130.0mm,angle=0.0]{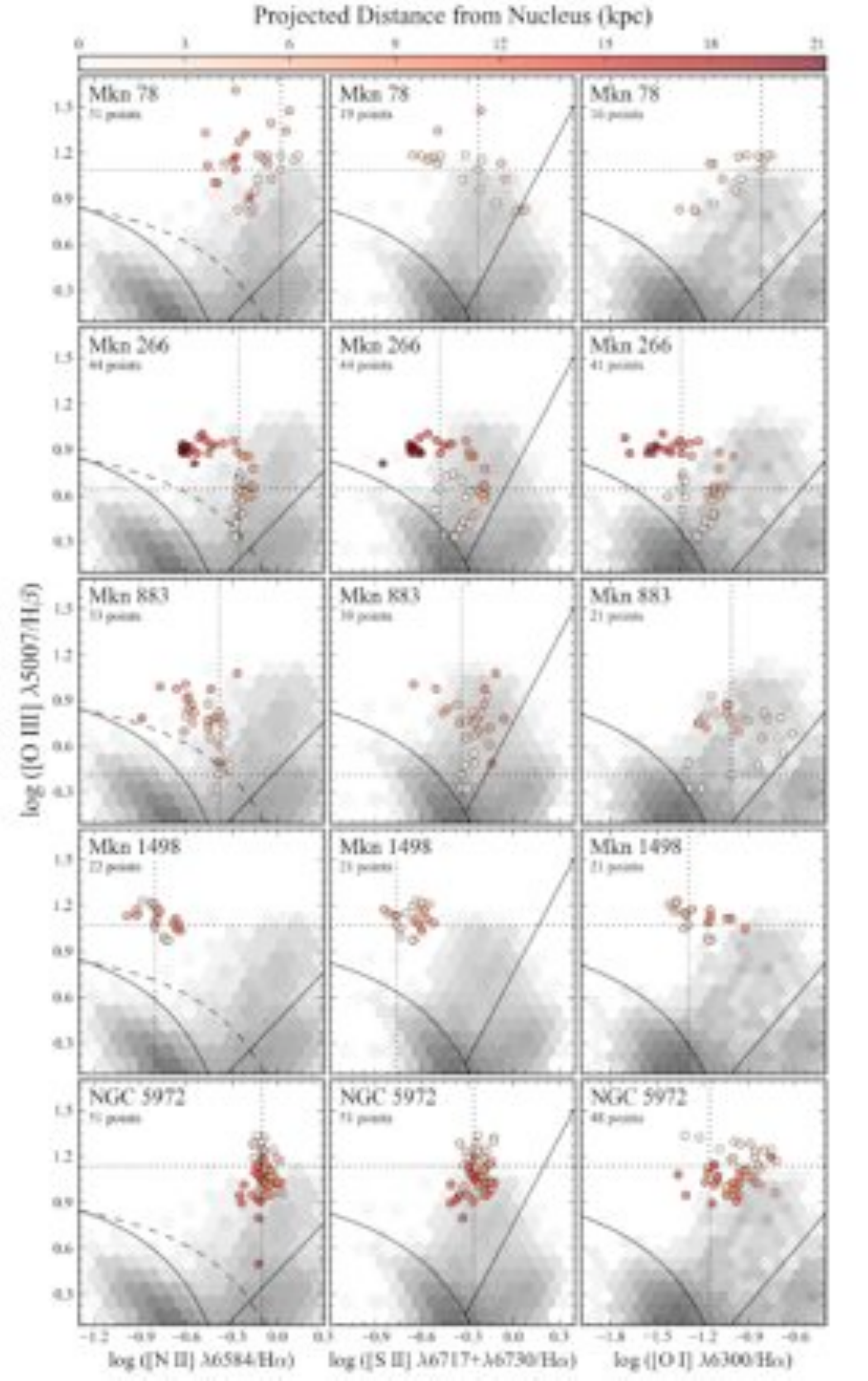} 
\caption{Baldwin-Phillips-Terlevich (BPT) diagrams for spatial slices in Lick data. Each row shows the three classical BPT diagrams for each object, highlighting radial ionization behaviors. The nucleus is indicated by crosshairs, with colors changing from white to red with increasing distance from the nucleus. The greyscale background and dividing lines are the same as in Figure 4; these show only the region around the AGN loci in each case for discrimination of detail. Nearly all measurements lie firmly in the AGN domain, with possible exceptions in some regions of SDSS 1005 and SDSS 0955. The starburst/AGN
ionization boundary from Kewley et al. (2001) is shown as the red full curve, while the boundary from
Kauffmann et al. (2003) is the black dashed curve.
All these measurements lie firmly in the AGN domain, with the possible exception of two regions in SDSS 1005
and the nucleus of Mkn 883, whose
red spectrum shows a broad-line region and strong {[O I]}. The greyscale 
show the density of points representing low-redshift galactic nuclei 
in the SDSS, from Schawinski et al. (2010b), which we
also follow in adopting the straight line as the LINER/Seyfert boundary}
 \label{fig-BPTplots}
\end{figure*} 

\begin{figure*} 
\includegraphics[width=150.mm,angle=0]{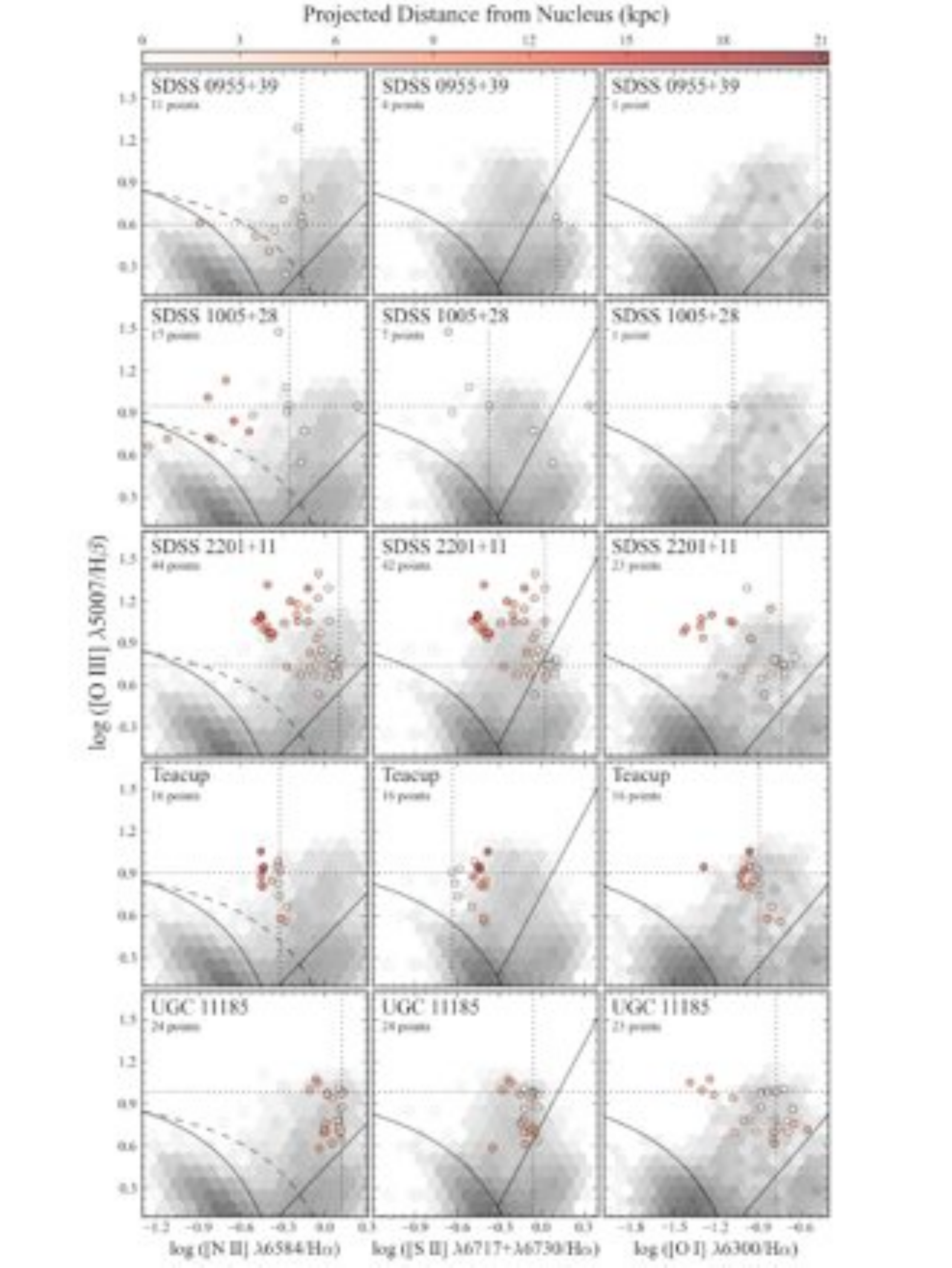} 
\contcaption{} 
\label{fig-velocity2} 
\end{figure*}

\begin{figure*} 
\includegraphics[width=160.0mm,angle=0.0]{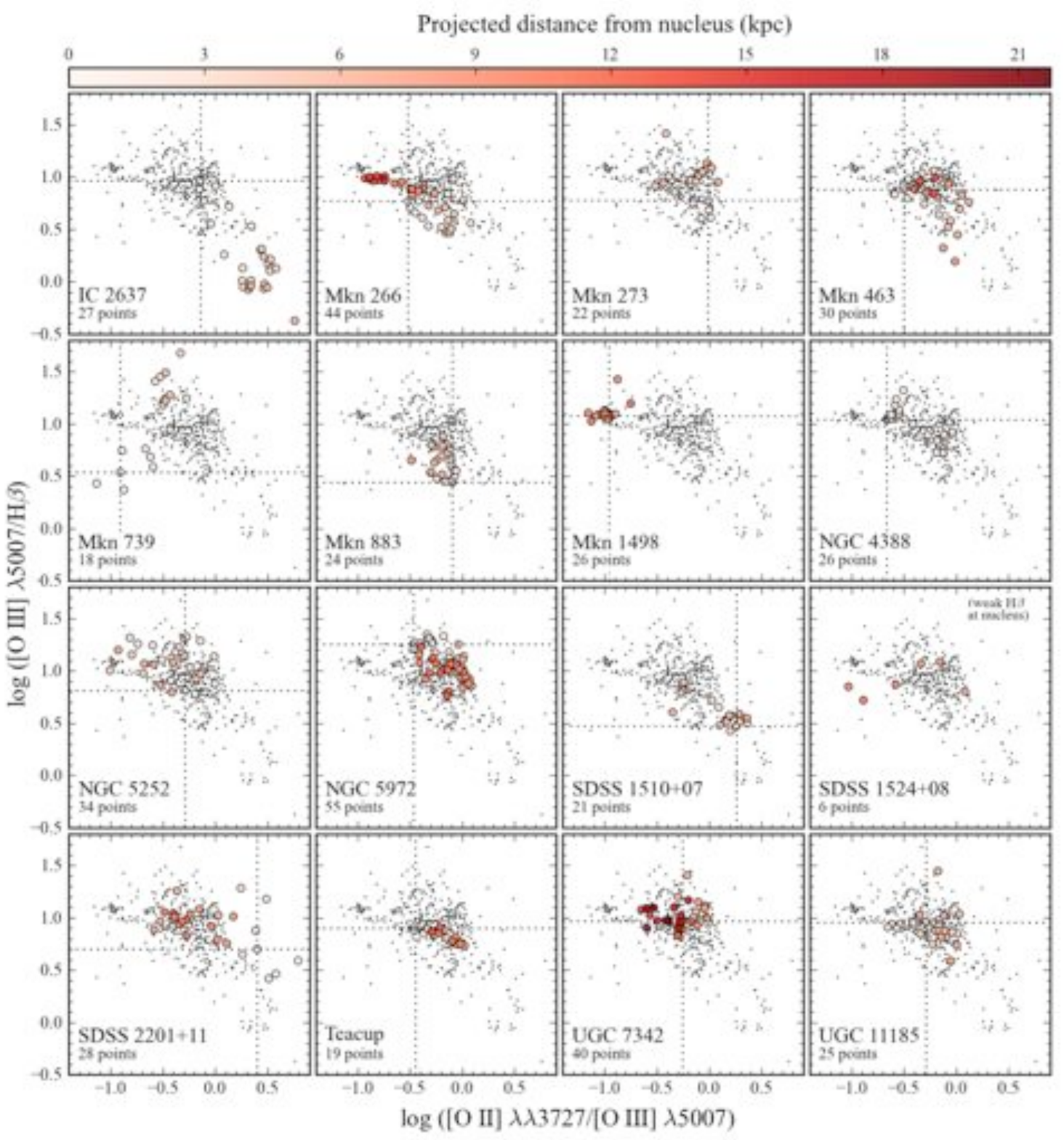} 
\caption{An alternate Baldwin-Phillips-Terlevich (BPT) diagram, usable when only blue spectra
are available (in this case, from the Kitt Peak GoldCam). As in Fig. \ref{fig-BPTplots}, points are colour-coded with
projected distance from the galaxy nuclei. Since Mkn 739 has a significant broad-line region, its H$\beta$ components
were deblended using a broad Gaussian and narrow Lorentzian, with narrow-line ratios plotted here.}
 \label{fig-bluebptplot}
\end{figure*}

\begin{figure} 
\includegraphics[width=85.mm,angle=0]{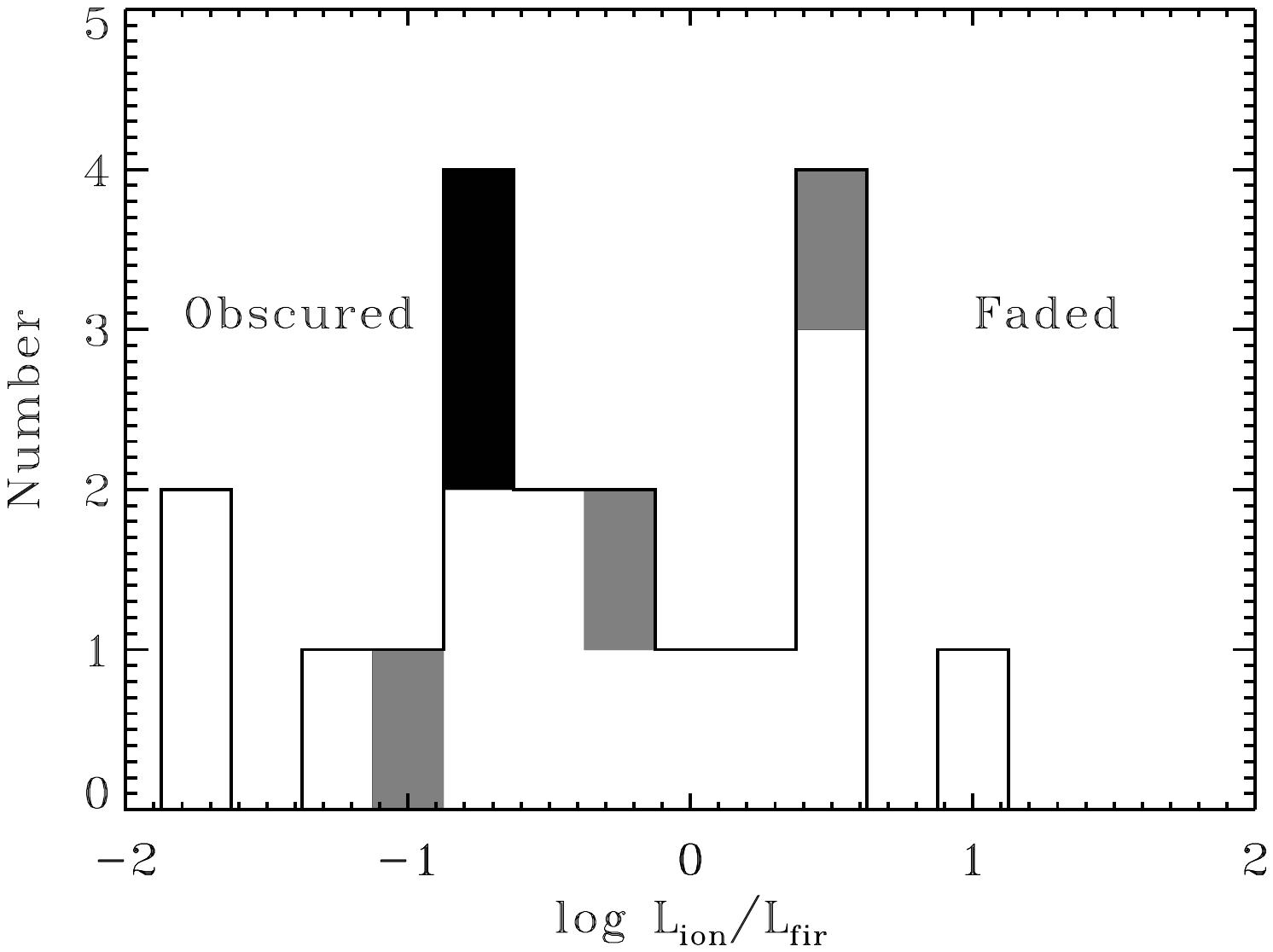} 
\caption{Distribution of the ratio $L_{ion}/L_{FIR}$ in the log. Small values are consistent with AGN heavily obscured along
our line of sight but not toward the emission-line clouds, while large values indicate fading of the AGN. Shaded regions
show that there is no obvious differences in the distributions among type 1 Seyferts (black), intermediate type 1.5 and 19 objets grey), and
type 2 nucle (white).}
 \label{fig-ratiohist}
\end{figure}

We use far-IR data to estimate (or limit, for nondetections) the amount of AGN radiation absorbed
(and reradiated) by nearby dense material, whether in an AGN ``torus" or in the
inner parts of the host interstellar medium. 
The FIR luminosity is conservatively high as an estimate of the potential
obscured AGN luminosity, since there may be a nontrivial 
contribution from star formation in the host galaxy as well as the AGN, and in some cases
companion galaxies might blend with the target in the FIR beam.
In a simple picture where a fraction $f$ of the AGN radiation
is absorbed by nearby dust and reradiated, the FIR luminosity will be
of order $L_{ion} f/4 \pi$, with an additional scaling factor of a few to account for
non-ionizing radiation heating the grains (which we omit at this point for the sake of a conservative calculation). 
For convenience, we approximate
the total far-IR output by the FIR parameter introduced for
Infrared Astronomical Satellite (IRAS) point-source
catalog data \citep{IRASgal}, a linear combination of flux values in the 60 and 100$\mu$ bands
which gives a reasonable approximation to the total flux between 42-122$\mu$. Numerically,
$$ {\rm FIR} ({\rm W  ~ m}^{-2}) = 1.26 \times 10^{-14} (2.58  f_{60} +  f_{100}) $$
for {\it IRAS} fluxes in the 60 and 100 $\mu$ bands given in Jy (multiplied by $10^7$ for a result
in ergs cm$^{-2}$ s$^{-1}$).
{\it IRAS} data were supplemented, where possible, by {\it Akari} data (\citealt{Akari}, 
\citealt{AkariFIS}, \citealt{AkariBSC}) of
higher accuracy. The positions of all these galaxies were covered in the {\it IRAS} survey, so we can assign typical
upper limits to nondetections depending on ecliptic latitude; {\it Akari} added two additional
detections not found in the IRAS data, using only quality 1 (confirmed detection) fluxes. 
For non-ULIRG objects ($L_{FIR} < 10^{45}$ erg s$^{-1}$), we can reproduce the {\it IRAS} FIR
parameter from {\it Akari} 90$\mu$ fluxes and mean colours via
$$ {\rm FIR} ({\rm W  ~ m}^{-2}) \approx 5.0 \times 10^{-14} f_{90}$$ with 30\% accuracy ($\pm 0.15$ dex), and we use this
to fill in FIR luminosities for the objects detected only by {\it Akari}. 
The input values and results of this energy-balance test are shown in Table \ref{tbl-energy}.
Within our sample, Mkn 273 and Mkn 266 are classic ultraluminous infrared galaxies (ULIRGs),
with $L_{FIR} = 1-5 \times 10^{45}$ erg s$^{-1}$. Some of the others remain undetected in 
both the IRAS and {\it Akari} surveys, leading to limits typically $< 10^{44}$ erg s$^{-1}$.
An index of whether the extended clouds can be ionized by an obscured AGN is provided by the
ratio of required ionizing luminosity to FIR luminosity, tabulated in Table \ref{tbl=-luminosity}. These values are all
lower limits, since the ionizing luminosity is a lower limit
This ratio ranges from 0.02 to values $>12$ (FIg. \ref{fig-ratiohist}). Low values clearly represent AGN
which are strongly obscured along our line of sight but not toward the EELR clouds. Large
ratios indicate long-term fading of the AGN, a spectral shape strongly peaked in the
ionizing UV, or very specific geometry for obscuring material, and thus indicate objects
worthy of close attention.

Arguments for long-timescale variations in the central sources here depend on our having
 estimates for their total luminosity as seen directly, which could in principle fail either if
 their ionizing radiation were collimated by something other than obscuration, or the
 spectral shapes in the deep ultraviolet differ from our expectations based on the UV
 and X-ray behavior of familiar AGN. Collimation by relativistic beaming would not account for the
 combinations of opening angle and flux ratio required (as found for Hanny's Voorwerp; Keel et al. in preparation).
 A spectral solution to the behavior would require an extreme-ultraviolet 'bump" dominating the ionizing flux
 from 13.6-54 eV by  more than an order of magnitude. However, known AGN do not provide 
 evident examples of either solution; the most straightforward interpretation of the data suggests that some of these clouds are ionized by AGN
 which have faded  over the differential light-travel time between our views of the clouds and nuclei.

\section{Nuclear and extended radio emission}

To further characterize the AGN in these galaxies, we collected radio fluxes at 1.4 GHz from the National Radio Astronomy Observatory Very 
Large Array Sky Survey
(NVSS) source catalog
\citep{NVSS}. All but two objects (SDSS 1510+07 and 1005+28) were detected
above the 2.5 mJy survey limit;
the source luminosity L(1.4 GHz) ranges from $ <1.3 \times 10^{22}$ 
W Hz$^{-1}$ to
$2.0 \times 10^{24}$, the latter for the double source associated
with NGC 5972 and comprising 94\% of the galaxy's total flux. Eight of the galaxies
qualify as radio-loud if one uses a simple, representative division at
$10^{23}$ W Hz$^{-1}$, and only one lies above 10$^{24}$. This
one - NGC 5972 - represents an interesting departure from the usual alignment of
emission-line and radio structure (section 7).

\section{Host and cloud morphology}

\begin{figure*} 
\includegraphics[width=160.mm,angle=0]{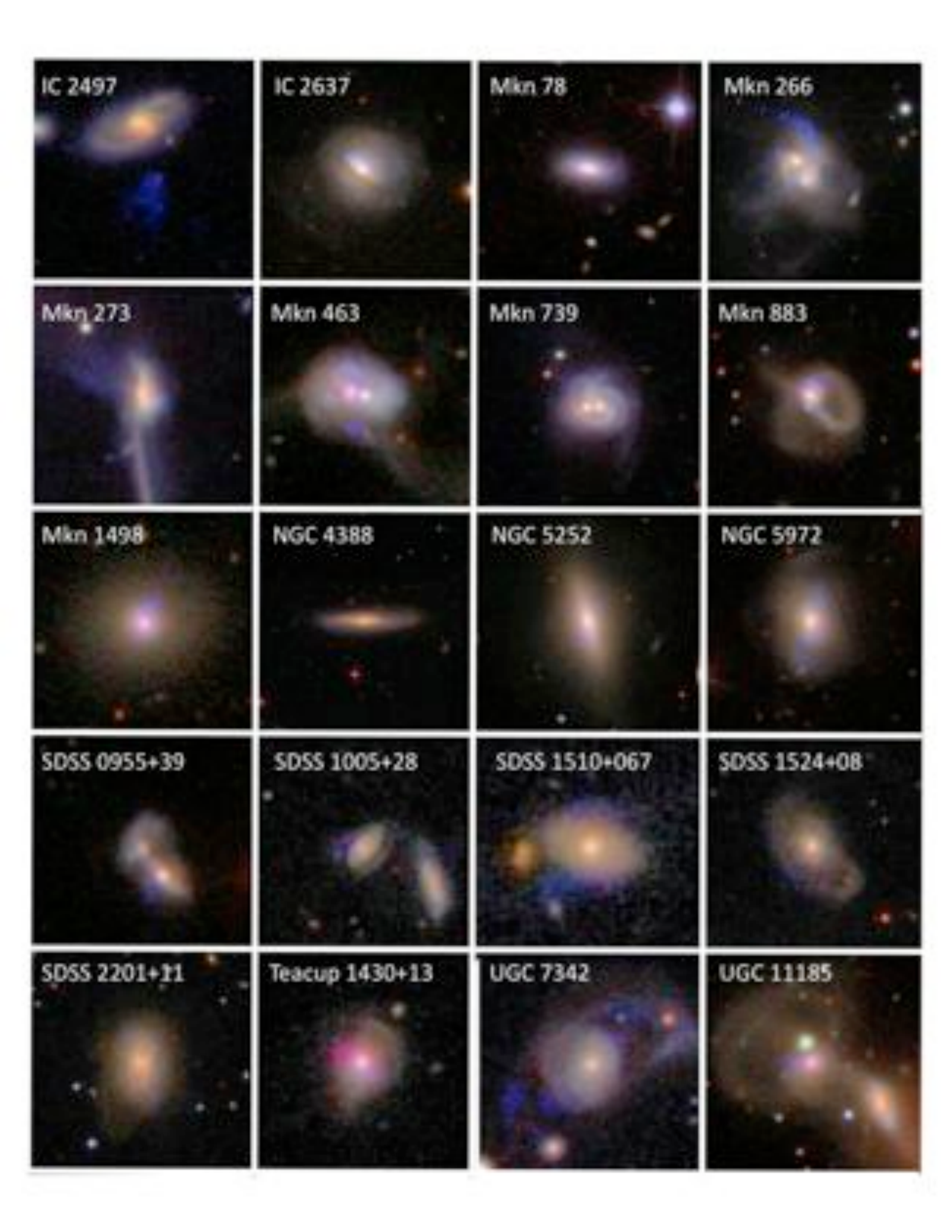} 
\caption{Host morphologies and ionized-gas structures in the Galaxy Zoo sample of
AGN-ionized clouds. These are based SDSS colour image products, with $gri$ filters mapped
to BGR for display. However, to enhance visibility of the ionized clouds, the contrast of
the $g$ filter has been increased in this figure, across the intensity ranges needed to show the gas effectively.
Each image cutout spans $70 \times 70$ kpc with north at the top. As in Table
\ref{tbl-morphology}, IC 2497 is included for comparison.}
 \label{fig-montage}
\end{figure*}

The examples of Hanny's Voorwerp \citep{Josza2009} and NGC 5252 \citep{Prieto}
suggest that a common source of extraplanar gas at large radii
is tidal debris. The host morphologies of the galaxies where
we find extended ionized clouds support this notion. Table 
\ref{tbl-morphology} lists morphological information on these
galaxies, including warps, close companions, asymmetries, or ongoing mergers. The actual incidence of tidal structures will be higher - for 
example, the inclined ring of gas with distinct kinematics in
NGC 5252 has no optical counterpart. Of the 19 confirmed large-scale clouds, 14 are in systems classified 
from SDSS data alone as interacting, merging, or postmerger (still showing tidal tails). This remarkably high 
incidence of disturbed systems (at least 73\%, even without including NGC 5252) supports the idea that most very extended 
emission-line clouds around local AGN represent illuminated tidal debris. We illustrate this in Fig.
\ref{fig-montage}, showing the SDSS colour images with the $g$-band  [O III] contribution enhanced to show the
clouds' locations. In this section, we include IC 2497/Hanny's Voorwerp in the tabulations for comparison.
A striking instance of a QSO ionizing gas in a companion and tidal tail, on similar scales $\approx 40$ kpc,
has been reported by \cite{daSilva}.

Several of these galaxies show discs seen nearly edge-on. From these, it is clear that the ionizing radiation
can emerge well away from the disc poles. The projected angles from stellar disc to the axis of the ionized
clouds, when it is well defined, range from 30--54$^\circ$. This fits with the statistics reported by \cite{Schmitt1997},
in which obscured (type 2) objects show a wider range of angle than type 1 objects between the host-galaxy axes and the AGN axes as
traced by radio jets.

We see both one- and two-sided emission regions. The two-sided regions are generally highly symmetric in angular
location, although not necessarily in radial extent or surface brightness,
fitting with biconical illumination patterns. As listed in Table \ref{tbl-morphology}, 9 of 19 of our confirmed
objects have emission detected on both sides of the nucleus. Particularly in very disturbed systems, a strong asymmetry may
reflect the location of cold gas rather than the pattern of escaping ionizing radiation, so that we cannot necessarily conclude that the
one-sided clouds are in galaxies that do not have two-sided radiation patterns

\begin{figure} 
\includegraphics[width=87.mm,angle=0]{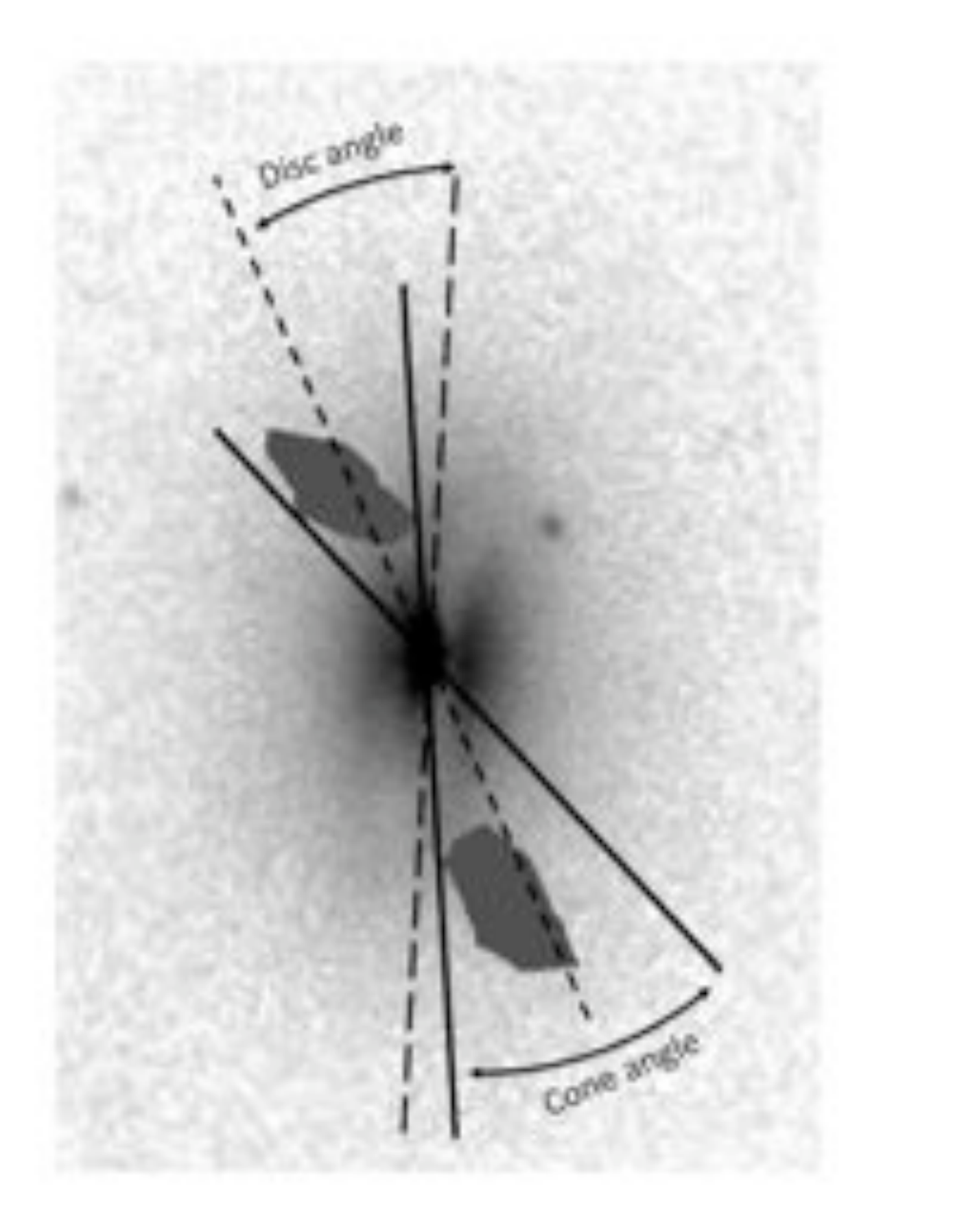} 
\caption{Illustration of angles listed in Table \ref{tbl-morphology}. The disc angle, defined only for
highly inclined and reasonably symmetric host disks, is measured from the projected major axis of the disc
to the midpoint of the cloud (bi)cones. The cone angle, as shown, measures the width of the paired triangles
about the nucleus that encompass the extranuclear emission detected in available images. The base image
is the SDSS $r$ observation of SDSS 2201+11, with the [O III] clouds shown schematically.}
 \label{fig-angles}
\end{figure}

The angular width of regions of escaping radiation may constrain the geometry of obscuring regions, if the ionization is bounded
by the availability of radiation rather than gas. We list, in Table \ref{tbl-morphology}, a cone angle, which is the projected angular 
width of each half of a notional bicone encompassing all the high-ionization regions seen in our images, outside of a usual
nuclear emission region (Fig. \ref{fig-angles}). Projection effects make the observed angle an upper limit to the three-dimensional opening angle of
a cone. The sample of large emission clouds exhibits a wide range, from 23--112$^\circ$. The narrower ones are challenging
to understand from obscuration by a circumnuclear torus, suggesting absorbers that are geometrically quite thick compared to the opening angle
for escape of ionizing radiation, to an extent which might better be described as an obscuring shell with small polar holes. However, some of these 
objects have dual clouds in very symmetric locations, which would be most naturally explained by such a scheme.

Several of the two-sided clouds show near-symmetry in radial extent on opposite sides of the galaxy. This could reflect
episodic activity on the light-travel time scale, although front-back geometric effects would generally break an exact symmetry.

\section{Kinematics of ionized gas}

Extended ionized regions around AGN may commonly be separated into kinematically quiescent components,
such as would be seen for disk gas ionized from the nucleus, and outflow, with additional superimposed radial
motion which might be manifested in a well-sampled velocity field as misalignment of the velocity field with
the galaxy morphology if the superimposed velocity components are not spectrally resoved \citep{Barbosa}. In addition, for 
disturbed systems, tidal features may show motions decoupled from the disk itself. We consider here the information on gas kinematics provided
by our spectra, noting that in most cases we sample only a single position angle through each galaxy.

Redshifts were measured for each pixel along the slit using Gaussian fitting in IRAF. We show results for
[O III], H$\beta$, and when observed,  H$\alpha$ and [N II].
Velocity errors are based on propagation of photon statistics \citep{Keel1996}.
		
Fig. \ref{fig-velocity1} shows a selection of these velocity slices, relative to the nucleus in each case.
Despite the angular offset from the edge-on disk, the gas velocities in SDSS 2201+11 are continuous
with the pattern in the inner rotating section, and closely symmetric. Similarly, the emission
clouds in NGC 5972 fall along an extrapolation of the inner-disk rotation curve (as traced by
[O III]). Despite its very disturbed morphology, UGC 11185 shows near-symmetry in kinematics,
with a very strong velocity gradient crossing the nucleus.

Other systems in our sample show less ordered velocity slices. The gas in UGC 7342 at all radii has a single
sense of motion on each side of the nucleus, but local departures have amplitudes up to 120 km s$^{-1}$.
A central gradient in the Teacup (SDSS 1430+13) may be reversed where the slit crosses its loop of emission.
The kinematics in Mkn 883 and Mkn 739 are very disordered, as expected for a merging system. In Mkn 78,
multiple components are seen in the inner few arcseconds, even in [Ne V]  \citep{Fischer}.

The northern filament in Mkn 266 presents interesting kinematic behavior, with a large and consistent
velocity offset between [O III] and H$\alpha$, H$\beta$. This difference is seen in spectra from both spectrographs.
A likely explanation is superposition of structures with quite different ionization states as well as velocity, so that the
weighting of lines in our spectra, even though they are not separately resolved, gives different velocity centroids. 
The offset is close to 50 km s$^{-1}$ along the entire filament. Localized instances of similar
mismatches between [O III] and H$\beta$ velocities on one side of SDSS 2201+11 and possibly in NCG 5972. 
Outflows are typically inferred from blue wings on [O III], but far from the nuclei where disk
extinction is unlikely to be a major effect; outflows could produce relative redshifts or blueshifts. One corollary
of this distinction is that there exists a gas component of much higher excitation than implied by 
the ionization parameter we derived from
ratios of total line flux at these locations, suggesting higher ionizing luminosities in these galaxies.

\begin{figure*} 
\includegraphics[width=155.mm,angle=0]{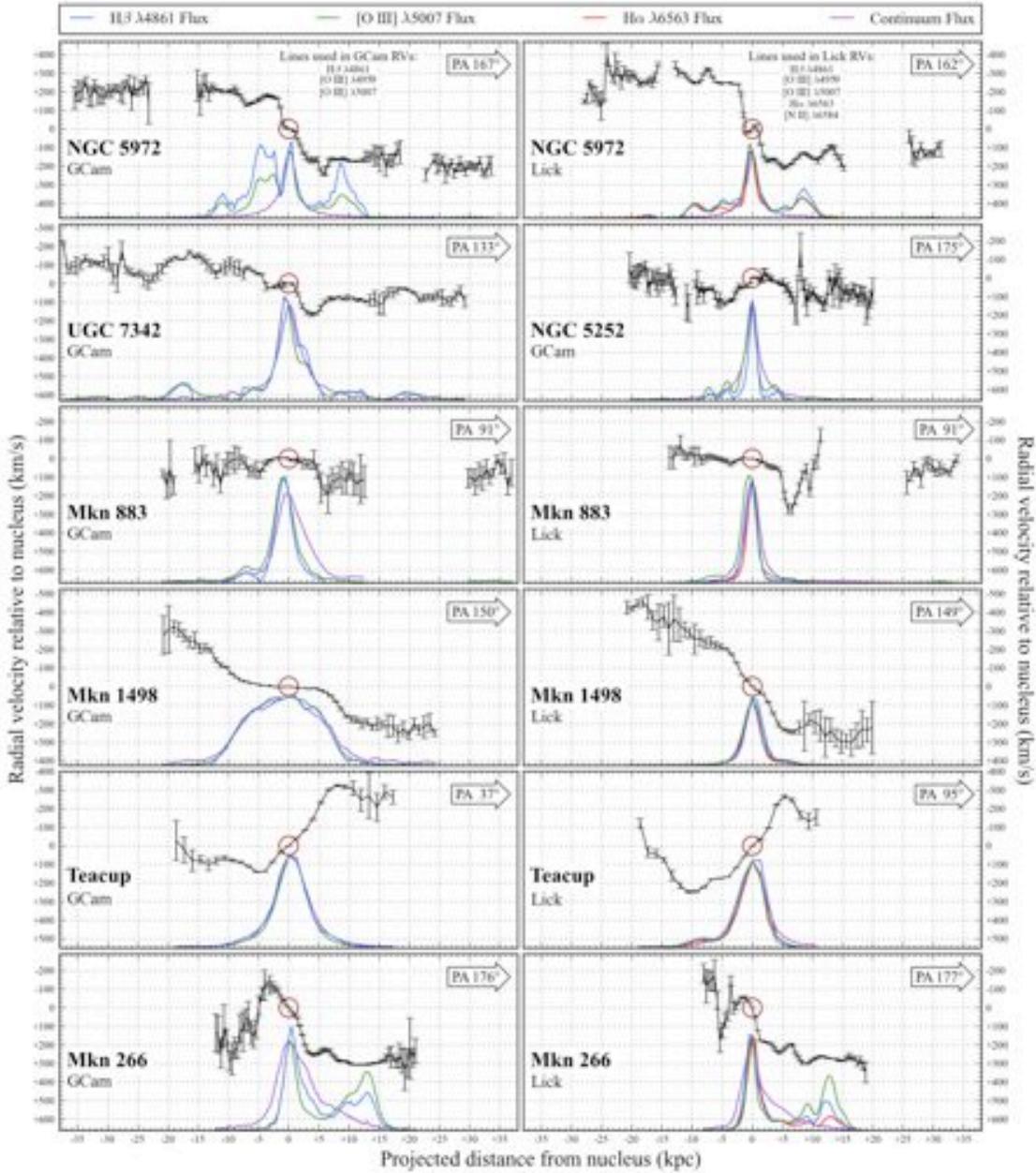} 
\caption{Radial-velocity slices (shown relative to the nuclei) along our slit locations for selected hosts and AGN clouds.
Nuclei are indicated by dark red circles. Lines used in calculating radial velocites from GCam and Lick data sets are indicated in the plots for NGC 5972, one of several objects observed at both locations though typically at slightly different position angles.
In some cases,
the extended emission follows the inner rotation closely even for extraplanar gas, as in SDSS 2201+11
and NGC 5972. Others are more chaotic, as expected for mergers. Particularly for Mkn 266, some structures
show significant differences between the lines, indicating that we are seeing blends of multiple
components with very different line ratios. Intensity slices in the continuum and lines are plotted across the
bottom to show correlations between location and velocity structure, scaled in flux to match [O III] $\lambda$ . Velocity errors are calculated as $\pm 2 \sigma$ 
from photon statistics following Keel (1996); in some cases, larger errors may be appropriate from such
factors as blending of multiple components.}
 \label{fig-velocity1}
\end{figure*} 

 \begin{figure*} 
\includegraphics[width=155.mm,angle=0]{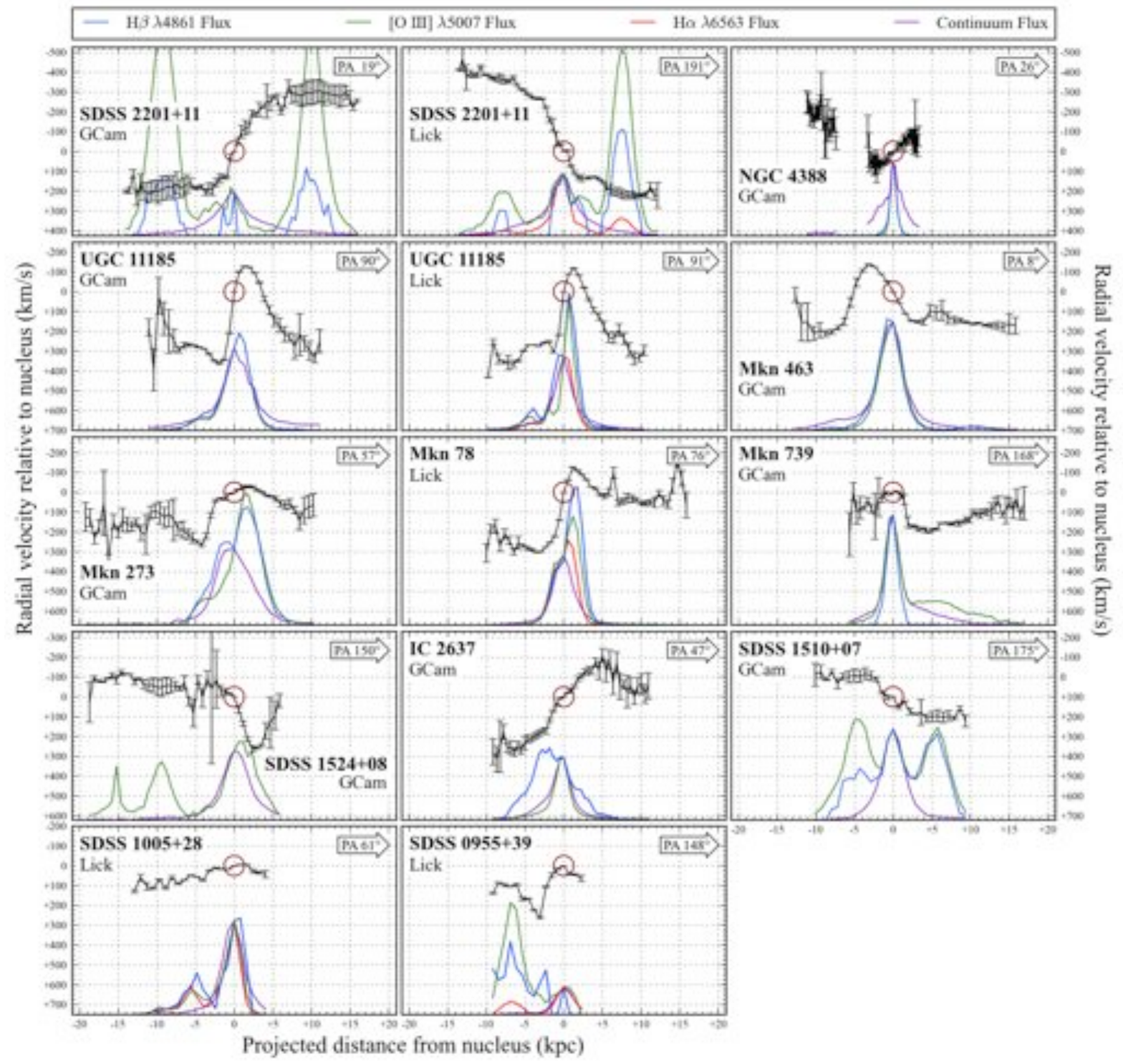} 
\contcaption{} 
\label{fig-velocity2} 
\end{figure*}

The relatively quiescent kinematics of most of these features may indicate a contrast in origin of the
extended gas when compared with the radio-loud QSOs  (\citealt{FuStockton1}, \citealt{FuStockton2}). In their sample,
modest star-formation rates led them to suggest that much of the ionized gas was expelled from the system by
the launch of powerful radio jets. The galaxies in our EELR sample are mostly radio-quiet (or at least radio-weak), as noted above. Also
unlike their QSO sample, we see significant metallicity differences between the nuclei and EELRs, most
strongly shown in the [N II]/H$\alpha$ ratio, again consistent with the EELR gas having an external tidal origin.

\section{Noteworthy systems}

From our results or previously reported data, several of these galaxies have interesting individual features.

The inner parts of the EELR in Mkn 78 have long been known (\citealt{Pedlar1989}, who detected much of the [O III] extent we observe), 
and observed with HST in both imaging and
spectroscopic modes (\citealt{Capetti1994}, \citealt{WW2004}). A detailed fit to the HST radial velocities explains the double line profiles near the core
without requiring a second AGN \citep{Fischer}; optical and near-IR line spectra suggest that the gas is photoionized from the nucleus with
at most a very localized role for
excitation by interaction with small-scale radio jets (\citealt{Whittle2005}, \citealt{Almeida}). Our data also show the complex spatial and velocity structure
in the inner few arcseconds. The outer emission is spatially smooth, and is measured to much larger radii
in our spectra than in the initial SDSS imaging detection.

In Mkn 883, the blue line ratios indicate that it lies near the Seyfert/starburst boundary. Only in the red
do weak broad H$\alpha$ and [O I] definitely indicate an AGN. We do not detect a broad component at H$\beta$.

For NGC 4388, the SDSS images detect only a few inner knots of the extensive emission region revealed by, for example, {\it Subaru} imaging
\cite{Yoshida}.
Our spectra detect more of this structure. Our cone angle is estimated from the {\it Subaru} image. Detailed spectroscopy by
\cite{Yoshida2004} confirms that this distant gas is photoionized by the AGN.

NGC 5252 has been considered the archetypal Seyfert galaxy with ionization cones. The implied energetics of the nucleus depend critically on
the density in the ionized filaments. Our implied limits from photoionization balance via surface brightness in H$\beta$ are significantly greater than
the values suggested from pressure balance with the galaxy's hot ISM \citep{Dadina}, while we concur with the X-ray results that the
ionization parameter remains roughly constant with radius among the ionized features. In turn, the interpretation of this behavior
depends on fine structure (much still unresolved) in the emission-line filaments, as seen in Hanny's Voorwerp \citep{Keel2011}.

Two objects in this sample appear to violate the usual pattern of ionization cones encompassing radio-source axes. NGC 5972 is the 
most radio-luminous of our galaxies, and shows
a typical double-lobed structure \cite{Condon88}. The 
lobes are separated by 9.4' (330 kpc) in projection, and 
are oriented near PA $100^\circ$, quite different from the optical 
emission at PA $167^\circ$. In this source, the most radio-powerful 
in our sample, the very different geometries of the line and radio emission
make ionization from interaction with the radio plasma unlikely, and their near-perpendicular orientation
is unlike the typical case for Seyfert galaxies
\citep{Wilson1994}.
This could be explained if the ionization cones have extremely broad opening angles,
or if the radio structure makes a dramatic and yet-unobserved twist on small scales.
Similarly, Mkn 1498 is associated with a giant low-frequency double radio source \cite{Rottgering}, with
projected separation 1.1 Mpc. In this case as well, the orientations of the emission-line structures and
the large radio source differ strongly, by about $70^\circ$.

SDSS 1430+13, the ``Teacup" AGN, is distinguished by a 5-kpc loop of ionized gas. The Faint Images of the Radio Sky at Twenty-one centimeters (FIRST) Very 
Large Array
data at 20 cm show extended structure roughly coextensive with this feature, possibly indicating a related origin.

In both SDSS 2201+11 and SDSS 1111-00 (the latter observed spectroscopically but with emission smaller
than our 10-kpc limit), the extranuclear clouds outshine the AGN itself in the [O III] lines.
 
UGC 11185 shows a second, weaker set of emission-line components near the nucleus, peaking about 1.8" to the east along our slit, roughly 600 km s$^{-1}$ to the red of the main peaks, and including about 1/4 of the nuclear [O III] flux within a $2 \times 3$" aperture.

In both Mkn 463 and Mkn 739, {\it Chandra} imaging has furnished evidence for double AGN (\citealt{Koss}, \citealt{Bianchi}). In both cases,
the emission regions are much larger than the separation between AGN components, so we do not know whether the ionization is
associated mostly with one or the other. More detailed [O III] images could resolve this. However the flux sources are
apportioned between components, one AGN in each system must have an ionizing/FIR ratio at least as high as our tabulated limit.
Several earlier studies have noted the extended [O III] emission around Mkn 463 (\citealt{Mazzarella}, \citealt{Uomoto}, \citealt{
Chatzichristou}).

\cite{Wu2011} summarize polarimetric detections of ``hidden" broad-line regions in nearby AGN.
Their list includes four of the nuclei in our sample: NGC 4388, NGC 5252, Mkn 78, and Mkn 463.
Broad wings to the Balmer lines are seen in Mkn 266 (southwestern nucleus) and Mkn 739 (eastern nucleus), making them clear Sy 1 nuclei with ``non-hidden"
broad-line regions. Weak wings are seen at H$\beta$ in Mkn 1498, which would then be classified as a type 1.8 object
\citep{DEO77}.

\section{Conclusions}

Volunteers in the Galaxy Zoo project have carried out a search for AGN-ionized gas clouds on large scales
(10-40 kpc). This paper has documented the search, and spectroscopic observations of candidates yielding
19 such features. These clouds were classified as AGN-photoionized based on their locations in the
Baldwin-Phillips-Terlevich (BPT) line-ratio diagrams, strength of [Ne V] and He II emission, and (when
measurements are sufficient) modest electron temperatures $T_e < 20,000$ K, consistent with photoionization but not with shock heating.
Most of the host galaxies show signs of interaction, suggesting that the extended ionized gas in many cases rises from
tidal tails.

We consider upper and lower bounds to the luminosity of the AGN as it reached the clouds - lower limits from recombination and
upper limits from density and ionization parameter. We compare these with the obscured luminosity estimated from far-infrared
measurements; an excess in ionizing luminosity (or deficit in the far-IR) could signal long-term variability of the AGN.  The ratio
of ionizing to obscured luminosity spans a wide range, from 0.02 to $>12$. Over a third of them (7/19) exceed unity, making this kind of
energy deficit a common issue. Small values fit with an origin in
obscured AGN, requiring only a small fraction of the extreme ultraviolet to escape. In contrast, large values
may require a long-term fading of the AGN. An extreme case of this is represented by Hanny's Voorwerp near
IC 2497 \citep{Lintott2009}. In this object, the required ionizing flux indicates that the AGN has faded by a factor
$>1000$ within the last $\approx 2 \times 10^5$ years, sampling a timescale on which we otherwise have
no information. More detailed observations of this new sample, including pending X-ray measurements, could 
give statistics adequate to show how common such variations are.

An important use of this sample is in addressing the history of AGN luminosity - on what timescales do episodes of
high luminosity persist and fade? Broad arguments suggest that AGN episodes extend over spans comparable to
the duration of a galaxy merger (several 10$^8$ years), if statistics associating excess AGN with strong interactions and mergers are representative.
We note that establishing a link between galaxy interactions and AGN episodes has proven
remarkably elusive, with the results depending on details of comparison sample selection and what kind of AGN is studied; as recent
examples, \cite{Maia}, \cite{Alonso}, and \cite{Li} 
reached different conclusions - a null result, enhancement limited to certain kinds of AGN, or a weak
overall enhancement of AGN - from similar analysis of nearby galaxy samples. 
Therefore, even within such long timespans, we have little information on how episodic the accretion and associated luminous output
might be.

\begin{figure} 
\includegraphics[width=70.mm,angle=90]{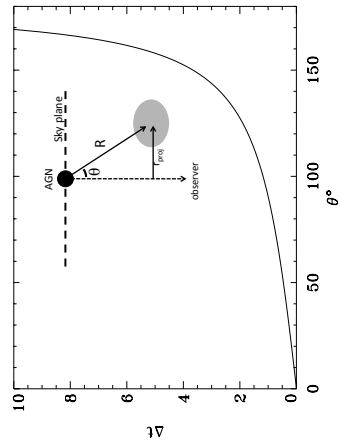} 
\caption{Behavior of the differential time delay $\Delta t$ between radiation seen directly from the nucleus and reprocessed through a
cloud at distance $R$ from the nucleus, as a function of angle $\theta$ from the line of sight, for fixed value of projected
radius $r_{proj}$. The inset diagram shows the geometric quantities. Units of $\Delta t$ are $r_{proj}/c.$ }
 \label{fig-echo}
\end{figure}

The relevant equation for time delay between radiation reaching us directly from the nucleus and that reprocessed in a cloud 
follow usage for light echos in ordinary reflection, except that here we are constrained by the location of
gas so we deal with a constant observed radius $r_{proj}$ and unknown angle $\theta$ between the illumination direction and the plane of the sky; and
the long recombination times at low density impose a convolution with a nontrivial time span for response. With the geometry
defined in Fig. \ref{fig-echo}, keeping $r_{proj}$ fixed by the observations means that the geometrical time delay $\Delta t$ for observing reprocessed nuclear
radiation depends on the the viewing angle $\theta$ (from observer to nucleus to the cloud, with a cloud along the line of sight
at zero and increasing away from the observer) as given by
$$  \Delta t = {{r_{proj}} \over {c \sin \theta}} (1 - \cos \theta)      \eqno{(1)}$$
derived in the approximation of infinite distance from the observer. Two-sided symmetric sets of clouds have progressively much
longer differential delays when seen with their axis near the line of sight, so that a faded source in this regime should eventually be seen ionizing only the farside cloud.
Our ability to reconstruct the actual distribution of $\theta$ is hampered by an inner cutoff in $r_{proj}$ (10 kpc, so that the cloud detection
is not hampered by galaxy starlight) and lack of knowledge of the distribution of cloud extent from the nuclei. To be conservative, our calculations
of ionizing luminosity (above) assume $\theta = 90^\circ$, the minimum possible distance for the nucleus and thus minimal ionizing luminosity,

We might expect our sample to be complete at least for the lowest redshifts and highest surface brightnesses, but there are a few objects
with selection priority as high as some of our cloud hosts for which we do not yet have confirming optical spectra. Of our 19 confirmations,
14 were found in both the targeted and serendipitous searches. Two were found only in the targeted search, and 3 in the serendipitous
search. Of these 3, one (SDSS J095559.88+395446.9) had no previous optical spectrum and could not have been included in the targeted sample. 

A first hint as to characteristic timescales comes for the relative numbers of galaxies with and without deficits in ionizing luminosity, since the
ones with deficits in the energy budget would be seen during the appropriate delay time after fading of the nucleus. There is no obvious
reason for this ratio to be biased in our sample, since the serendipitous survey was independent of the presence of an AGN, and even in the targeted
search there are many AGN which are too weak to ionize the extended gas; in essence, given a luminous AGN, our selection is for
objects with outlying gas available to be ionized. In a toy model where all objects' delay times $\Delta t$ are equal, the timescale for the AGN to be
at high luminosity before fading would be of order $\Delta t n_{bright}/ n_{faded}$.  From Table \ref{tbl-energy}, our estimate is $1.2 \Delta t$  when we divide the
bright and faded groups at an ionizing/FIR ratio of 1.5. The projected extent of the clouds $r_{max}$ from our [O III] data is listed in table
 \ref{tbl-morphology}; for the 19 galaxies in our sample, the mean value is 19 kpc with a median of 17. 
 For a typical projected extent 20 kpc, this becomes a range 25,000--175,000 years, taking the
sample to populate values of $\theta = 90 \pm 50^\circ$ at this small sample size. For the luminosity range of Seyfert galaxies we have probed,
the fading may be an order of magnitude in ionizing luminosity, but this sample includes no cases in which we see AGN-ionized clouds
around a galaxy with no optical trace of an AGN. IC 2497 \citep{Lintott2009} must be extreme in this respect, having faded from a QSO
to a borderline LINER./Sy 2 nucleus. As noted by \cite{Schawinski2010a}, these timescales are rapid compared to expectations from 
scaling up the behavior of accretion disks around stellar-mass black holes, perhaps indicating that disk self-gravity enhances the growth
of accretion instabilities.

There are several directions in which we can expand this study. In a ``Dead Quasar Survey", we are conducting [O III] imaging of samples of luminous 
AGN hosts and galaxies without AGN signatures, to seek fainter (and possibly older) clouds than can be detected from the
SDSS $g$ images. H I selection should help pick out objects with
tidal tails in suitable positions to be ionized at tens of kpc from the core; we are beginning with the \cite{Kuo} sample of Seyfert galaxies mapped in H I.
For the ``faded" galaxies in thes sample, we are pursuing {\it XMM-Newton} and {\it HST} observations to clarify the obscuration toward the nucleus, seek any signs of outflow-induced star formation as seen in Hanny's Voorwerp, and refine estimates of the ionizing luminosity through the highest recombination-line
surface brightness in the clouds.

\section*{Acknowledgements}
This work would not have been possible without the contributions of citizen scientists as part of the Galaxy Zoo project. 
We particularly thank Hannah Hutchins, Elizabeth Baeten, Massimo Mezzoprete, Elizabeth Siegel, Aida Berges, and users
voyager1682002 and Caro, who each examined all of the galaxies in the targeted AGN sample, and in addition Christian
Manteuffel, for assistance in compiling the list of SDSS AGN candidates. We are grateful to the following additional
Galaxy Zoo participants who contributed to the targeted AGN search:
Michael Aarons, Mark Ackland, AdrianusV, Aerial, alexob6, Daniela Alice,
Norvan Allen, Anderstp, AndrewM, angst, Anjinsan, ARCHEV, artemiit, aryamwojn,
astrobrainiac, astronomicom1, Markku Autio, Michelle Ayers, Elisabeth Baeten,
R. Balick, Michael Balzer, Michael Derek Barnett, Kirsten Barr, Barbara Ann Barrett,
David Bartlett, Coral Benham, Aida Berges, Mark Bernaldo, Chiheb Boussema,
Gwen Brogmus, Dave Browne,  buddyjesus, David Burt, cadou,
caliz83, Capella, Alice Carlsen, Caro, Jiri Cejka, Theodore J. Celaya Sr.,
chairstar, Bruno Chiaranti, citisue3, Nick Clarke, Ana Claudia, cloud9, clua, David N. Cook,
coral, Gemma Coughlin, Rob Cowhey, Penny Cox, Laurence Cuffe, cyprien,
DancesWithWords, Darren, DarthKeribo, Lloyd Daub, daveb, dave3,
david\_mbe, david\_nw, Michael C. Davis Jr., distel, Dobador, Shane Dobkins,
drawm, Juliette Dowle, Elizabeth Duff, Graham Dungworth, dxjerlubb,
dzd, Michael Easterly, echo, Alan Eggleston, Thomas Erickson, ErroneousBee,
Falconet, firejuggler, frisken, Gino, glyphon, GNB080, gordhaddow, Michael Gronceski,
grrower1, Michael Hand, Thomas Hardy, Hans Heilman, Steph Hill, Thomas Hobbs,
Rick Holtz, Rob Hounsell, hrutter, Mikko Huovinen, IC1101, ixzrtxp,
Nina Jansen, Alain Jaureguiberry, jayton, jczoehdo, jhyatt, David E. Johnson,
Steve Johnson, David James Jones, John Kelly, khwdfnwit, Pat Kieran, KillerSkaarj,
kiske1, knuid, kokdeblade, Anuradha Koratkar, Michi Kovacs, kzhndepnd
Marc Laidlaw, laihro01, landersonzych, Lily Lau WW, lawless, Bill Lawrence,
Kathleen Littlefield, Liz, Marc Lluell, Michael Lopez, lpspieler,
luigimx, Lzsp, Michael MacIsaac, Christine Macmillan, Katie Malik, Steve Malone,
mardo, Lelah Marie, Mark, Michael Marling, Stephanie R. Marsala, Mauro Marussi,
marxpmp, Mark McCormack, Rob Mellor, Massimo Mezzoprete, mgn, Michaelr1415,
MichaelRoberts, MichaelSangerTx, milk\-n\-cookies, miraculix250, Elspeth Mitchell,
Graham Mitchell, mlvgofjedxv, mothic, Mukund, mykyij, NGC3372, Julian Nicol,
Rick Nowell, nrbeuw, Richard Oram, orion, oswego9050, pbungaro, Alice Peachey,
Thomas Perraudin, Amanda Peters, Erica Pinto, plummerj, Jim Porter,
Steven Porter, Richard Proctor, ptkypxdh, randa, RandyC, Kim Reece, Jessica Reeder,
RelativisticDog2, Thomas Rickenbach, ripw, rjwarmv, rnjrchd, Michael Roberts,
RobinMiller, Jim Robinson, roborali, Rona, Geoff Roynon, Paul Rutten,
Rynnfox, S4CCG, Michael Salmon, salteV, Jeroen Sassen, second\_try, Matt Sellick,
sheba, Alice Sheppard, SianElderxyz, Nanne Sierkstra, Michael Simmons, SJPorter,
skepticdetective, Stephen Sliva, Mark Smith, Sophie378, spat, Maria Steinrueck,
stella13, stellar390, John H. Stewart, Doug Stork, sumoworm, superhouse,
tadaemdg, Auralee Tamison, Chet Thomas, thom\_2, Michael Thorpe, timchem,
torres, Trixie64, Ramon van der Hilst, Marcel Veillette, Rob H.B. Velthuis,
John Venables, Michael Viguet, vkhtmhfigou, Aileen Waite, David Walland,
wbybjbpv, weezerd, Mark Westover, Julia Wilkinson, Nat T. Winston III,
Windsmurf, wpubphx, xuhtjhc, xzxupfqjd, and Mairi Yates. We also thank the referee,
who caught a mistake in calculating light-delay times and helped make the discussion
more comprehensive. Jean Tate helped to untangle some issues of participant discovery order.

W.C. Keel acknowledges support from a Dean's Leadership Board faculty fellowship.
C. J. Lintott acknowledges funding from The Leverhulme Trust 
and the STFC Science in Society Program. 
Galaxy Zoo was made possible by funding from a Jim Gray Research Fund from Microsoft and The Leverhulme Trust. S. D. Chojnowski participated through the SARA
Research Experiences for Undergraduates program funded by the US National Science
Foundation. This research is based on observations with AKARI, a JAXA project with the 
participation of ESA. We thank the Lick Observatory staff for their assistance in obtaining
the data. Support for the work of KS was provided by NASA through Einstein Postdoctoral Fellowship grant number PF9-00069 issued by the Chandra X-ray Observatory Center, which is operated by the Smithsonian Astrophysical Observatory for and on behalf of NASA under contract NAS8-03060.

Funding for the creation and distribution of the SDSS Archive has been
provided by the Alfred P. Sloan Foundation, the Participating Institutions,
the National Aeronautics and Space Administration, the National Science
Foundation, the U.S. Department of Energy, the Japanese Monbukagakusho,
and the Max Planck Society. The SDSS Web site is http://www.sdss.org/. 
The SDSS is managed by the Astrophysical Research Consortium (ARC) for
the Participating Institutions. The Participating Institutions are The
University of Chicago, Fermilab, the Institute for Advanced Study, the Japan
Participation Group, The Johns Hopkins University, Los Alamos National
Laboratory, the Max-Planck-Institute for Astronomy (MPIA), the
Max-Planck-Institute for Astrophysics (MPA), New Mexico State
University, Princeton University, the United States Naval Observatory, and
the University of Washington.

This research has made use of the NASA/IPAC Extragalactic Database (NED),
which is operated by the Jet Propulsion Laboratory, Caltech, under
contract with the National Aeronautics and Space Administration.

\begin{table*}
 \centering
 \begin{minipage}{140mm}
  \caption{Candidate AGN with extended emission-line clouds}
  \begin{tabular}{@{}llccccl@{}}
  \hline
Coordinate name       &       SDSS ObjID      &    $z$ &  Nucleus &  Name/note      & Search & Posted by\\
 \hline
SDSS J005607.66+254804.7 & 587740589487030353 & 0.1530  & Sy 1 & purple haze  &    S & ElisabethB\\
SDSS J003507.44+004502.1 & 587731187281494175 & 0.1205 & LINER &  blue arc & S & scott L\\
SDSS J004527.06+004237.6 & 587731187282608299 & 0.1096 & Sy 1.5 &                  &    S & davidjamesjones\\      
SDSS J013037.75+131251.9 & 587724197207212176 & 0.0721 & Sy 2 & CGCG436-065 &  T \\
SDSS J014238.47+000514.7 & 588015509280587804 & 0.1459 & Sy 1 &                 &        S & Tsering\\
SDSS J014644.82-004043.1 & 588015508207304746 & 0.0827 & Sy 1  &                 &       S & davidjamesjones\\
SDSS J030526.96+005144.9 & 588015510363373793 & 0.1181 & Sy 1 &               &        S & Mukund Vedapudi\\
SDSS J030639.58+000343.2 & 588015509289762862 & 0.1074 & Sy 1   &             &      S & mitch\\
SDSS J033013.26-053235.9 & 587724242842026028 & 0.0131 & Sy 1 &  NGC 1346    &    S & Half65\\    
SDSS J040548.78-061925.7 & 587727178476093634 & 0.0556  & Sy 1  &               &    S & echo-lily-mai \\
SDSS J074241.70+651037.8 & 758878270293868614 & 0.0371 & Sy 2  & Mkn 78     &  T \\
SDSS J075910.44+115156.7 & 588023046395527377 & 0.0503 & Sy 1   &              &      S & silverhaze\\
SDSS J080452.73+212050.2 & 588016878287650850 & 0.1242 & Sy 1& purple haze & S   & davidjamesjones\\
SDSS J082034.78+153111.3 & 587741532229337219 & 0.1435 & Sy 1 &                   &    S & Half65\\
SDSS J082342.37+482754.4 & 587725470667833620 & 0.0935 & Sy 2  &                  &    S & spiralmania\\
SDSS J082642.63+111555.5 &  587745244691628370 & 0.0884 & trans &            &   T \\
SDSS J083525.51+104925.7 & 587744873714679862 & 0.1172 & Sy 1 &                 &      S & spiralmania\\
SDSS J083818.43+333441.3 & 587732470387703859 & 0.0621 & LINER & KUG   &    T \\
SDSS J084002.36+294902.6 & 587735240637284507 & 0.0648 & Sy 2 & 4C 29.30  &  T \\
SDSS J084344.98+354942.0 & 587732484342415393 & 0.0539 & Sy 2  &                &   S T & laihro\\
SDSS J084518.51+142034.1 & 587742062124269645 & 0.0606 & Sy 1 & purple haze & S   & mitch\\ 
SDSS J084810.10+351534.0 & 587732470388818038 & 0.0573 & Sy 2 & KUG0845+354  & T \\
SDSS J084809.59+351530.3 & 587732470388818042 & 0.0567 & Sy 2 & KUG0845+354     & T \\
SDSS J084917.31+531755.7 & 587725470670585957 & 0.1112 & Sy 1 & purple haze &    S   & Tsering  \\
SDSS J085625.93+021310.5 & 587727944563884115 & 0.1251&  RG  &                 &   S & Bruno\\
SDSS J085729.84+064210.6 & 587734691423453446 & ---  & --- &                 &   S & lovethetropics\\
SDSS J085837.52+182221.5 & 587741708866289722 & 0.0588 & Sy 2 &           &      T \\
SDSS J085813.75+385631.8 & 587732053243855000 & 0.0884 & Sy 2  &         &    T \\
SDSS J090547.33+374738.2 & 587732152565432366 & 0.0475 & trans &             &  T \\
SDSS J090958.07+621450.4 & 587737826756198431 & 0.0261 & Sy 2  & NGC 2742A & S & Ioannab\\           
SDSS J091011.34+230717.9 & 587741421104136355 & 0.0362 & --- &                &      S    & Citizen\_Kirk     \\
SDSS J091708.26+292215.6 & 587738946130018402 & 0.0353 & Sy 2 &  KUG 0914+295 & S & Half65\\        
SDSS J093033.05+034443.6 & 587728880335257753 & 0.0911 & Sy 1 &                 &    S & ElisabethB \\
SDSS J094529.64-002154.7 & 587725074458345485 & 0.0515 & LINER &              &  T \\
SDSS J095559.88+395446.9 & 588016528244670522 & 0.0483 & --- &  violet plume  &   S & StephanieC\\
SDSS J100507.88+283038.6 & 587741392112451744 & 0.0517 & Sy 2  &                    &  S T & laihro\\
SDSS J100529.60+275844.2 & 587741391575580675 & 0.0555 & Sy 2   &                   &  S & mitch\\
SDSS J101128.26+260655.4 & 587741490365792357 & 0.1164  & Sy 2  &                    &  S & mitch\\
SDSS J101645.11+421025.5 & 588297863112294442 & 0.0553 & NLSy1 &               &  S  & mitch  \\
SDSS J102016.20+524756.9 & 587731499185602761 & 0.0689 & Sy 2  &                  &  S & \\
SDSS J102108.59+024058.5 & 587726033311498259 & --- &       &               &   S & mitch \\
SDSS J103734.22+140120.5 & 587735349101199404 & 0.2061 & Sy 2  &                  &  S & davidjamesjones \\
SDSS J104232.05+050241.9 & 587728880879992930 & 0.0271 & Sy 2 & NGC 3341 & S & mitch\\
SDSS J104326.47+110524.2 & 587734948595499096 & 0.0475 & Sy 1 & purple haze &   S & lovethetropics\\
SDSS J104515.28+421331.7 & 588017626147782828 & 0.0990 & LINER & bubble?     &  S & RandyC\\
SDSS J110157.90+101739.2 & 587732772658806856 & 0.0340 & Sy 1 &                      &  S & davidjamesjones\\
SDSS J110335.42+032014.7 & 587726033853022296 & 0.0531 & Sy 2    &                &    S & skwalker\\
SDSS J110445.46+041755.2 & 588010358543351857 & 0.0252 & Sy 2     &              &     S & ElisabethB\\
SDSS J110756.53+474434.8 & 588295840708755475 & 0.0727 & Sy 1     &             &      S & davidjamesjones\\
SDSS J111100.60-005334.9 & 588848898833580220 & 0.0904 & Sy 2    &                &  S T & ElisabethB\\
SDSS J111113.00+284242.7 & 587741532784361479 & 0.0294 & SB & NGC 3561A        &      T \\
SDSS J111113.18+284147.0 & 587741532784361477 & 0.0295 & Sy 2 & NGC 3561       &       T \\
SDSS J111349.74+093510.7 & 587734892748144649 & 0.0292  & Sy 1.5 & IC 2637        &   S   T & stellar190 \\
SDSS J111653.96+593146.8 & 587729387686461462 & 0.0815 & trans & VII Zw 384  & T & \\
SDSS J112534.58+523247.0 & 587732136456487055 & 0.0270 & SB  &                &       S & errattan \\
SDSS J112753.87+302138.6 & 587741491447070913 & 0.0736 & nonAGN&             &          S & paulrogers\\ 
SDSS J112942.51+235014.1 & 587742189363331247 & 0.1277 & LINER  &               &           S & ElisabethB\\
SDSS J113323.97+550415.8 & 587733081347063838 & 0.0085 & Sy 1 & Mkn 177 compn       & S & stellar190\\
SDSS J113629.36+213551.7 & 587742013279502427 & 0.0297 & Sy 1 & Mkn 739                  & S T & Budgieye\\
SDSS J113849.61+574243.4 & 587735696978215000 & 0.1162 & Sy 1  &                                &        S & ElisabethB\\
SDSS J114155.61+010516.7 & 588848901521277093 & 0.1365  & Sy 2 &                  &  S & lovethetropics\\
\hline
\end{tabular}
\end{minipage}
\label{tbl-candidates}
\end{table*}

\setcounter{table}{2}
\begin{table*}
\contcaption{}
 \centering
 \begin{minipage}{140mm}
  \caption{Candidate AGN with extended emission-line clouds}
  \begin{tabular}{@{}llccccl@{}}
  \hline
Coordinate name       &       SDSS ObjID      &    $z$ &  Nucleus &  Name/note      & Search & Posted by\\
 \hline
SDSS J114454.85+194635.3 & 588023669168537695 & 0.0274 & Sy 2  &                 & S     T & stellar190 \\
SDSS J114517.10+200121.8 & 588023669705474229 & 0.04953 & --- &   (possible)        &   S & Half65\\
SDSS J115140.70+675041.9 & 587725552285122567 & 0.0629 & Sy 2  &           &    S & StephanieC\\
SDSS J115739.07-023908.3 & 587724649256779921 & 0.1308 & Sy 1 & purple haze     &     S & c\_cld\\
SDSS J115906.89+101001.7 & 587732771591225359 & 0.1165 & Sy 1.8 & purple haze  &  S  & c\_cld\\ 
SDSS J120114.35-034041.0 & 587725039018311737 & 0.0196 & Sy 1 &  Mkn 1310          &   S & Milk\_n\_cookies\\
SDSS J120150.80+143323.9 & 587735348036370587 & 0.0677 & nonAGN &                &  S & DuffBeer\\
SDSS J120719.81+241155.8 & 587742189367066665 & 0.0505 & NLSy1 & purple haze Mkn 648 & S & davidjamesjones\\
SDSS J120939.43+643107.6 & 587729154134966352 & 0.1042  & Sy 2  &                &   S & mitch\\
SDSS J121418.25+293146.7 & 587741532253519916 & 0.0632 & Sy 2 &  Was 49ab          & S T & stellar190\\
SDSS J121431.32+402902.6 & 588017979429486656 & 0.1211& Sy  2  &                    &  S & mitch \\
SDSS J121452.41+591953.2 & 587729386079059975 & 0.0607 & Sy 2 &  VII Zw 444  &   S  & mitch\\     
SDSS J121553.08+051447.8 & 588010359624827047 & 0.0803 & LINER &                            &S  & codexluminati\\
SDSS J121819.30+291513.0 & 587739719750058064 & 0.0477 & Sy 2  & UGC 7342             &S T & stellar190\\
SDSS J122402.57+435814.0 & 588017603610935505 & 0.1040 & nonAGN &       &  S& fluffyporcupine \\
SDSS J122546.72+123942.7 & 588017566564155399 & 0.0086 & Sy 2 & NGC 4388                    & S T & RandyC\\
SDSS J122802.10+094347.9 & 587732771057434918 & 0.1534 & LINER  &                &    S & Bruno\\
SDSS J123034.25+033800.7 & 587726016682917948 & 0.01285 & LINER &                 & S    & StephanieC  \\
SDSS J123038.98+401614.4 & 587738947758456849 & 0.1322 & Sy 1.5 &  purple haze &  S & Tsering\\  
SDSS J123046.11+103317.3 & 587732772131504164 & 0.01540$^*$   & SB$^*$  & VPC 0764     & S & lovethetropics\\
SDSS J123113.12+120307.2 & 588017702933823557 & 0.1161  & Sy 1  &               &   S & lovethetropics\\
SDSS J124036.73+365004.3 & 587739096991334439 & 0.0404  & Sy 2  &                  &    S & stellar190\\
SDSS J124046.40+273353.5 & 587741602034090027 & 0.0565  & Sy 1.5  &             &    S & Tsering\\
SDSS J124103.66+273526.0 & 587741602034155555 & 0.2007  & Sy 1   &                &  S & Tsering\\
SDSS J124325.65+365525.3 & 587739096991596570 & 0.0839 & Sy 2 &                &    S & Bruno\\
SDSS J124450.84-042604.5 & 587745544806727722   &  0.0147$^*$   &  LINER$^*$ &   IC 0812   & S & Milk\_n\_Cookies\\
SDSS J124505.56+102433.2 & 587732772133011652 & 0.0976&  Sy 2  &                 &  S & davidjamesjones\\
SDSS J124511.84+230210.0 & 587742014897127434 & 0.02326  & --- &  IC 0813 &  S & elizabeth\\   
SDSS J125741.04+202347.7 & 588023670249750583 & 0.0807 & Sy 2 & IC 3929      & S T & c\_cld\\
SDSS J130007.06+183914.3 & 587742575372730410 & 0.1130 & Sy 2 &                     &  S & mitch\\
SDSS J130234.89+184122.3 & 588023668102856809 & 0.0656 & Sy 2 &                     & S & Mukund Vedapudi\\   
SDSS J130422.19+361543.1 & 587738950954385445 & 0.0443 & Sy 2 & WR 470        &   S & mitch\\
SDSS J130258.82+162427.7 & 587742773491531836 & 0.0673 & Sy 1 &  Mkn 783          &    S & stellar190 \\
SDSS J130509.98-033209.2 & 587725039025258588 & 0.0835 & Sy 1 & purple haze     &    S  & lovethetropics\\
SDSS J131555.15+212521.5 & 587742013289660465 & 0.0884 & Sy 1  &                 &    S T & davidjamesjones\\
SDSS J131639.74+445235.0 & 588017605762482225 & 0.0909 & Sy 1.9 &           &  S   & c\_cld  \\
SDSS J131913.93+132030.8 & 587736802936684556 & 0.0960 & Sy 2  &                  &    S & mitch\\
SDSS J132340.31-012749.1 & 587725041711644785 & 0.0767 & Sy 1 &         &  S & IC 1101 \\   
SDSS J132540.23+275146.1 & 587739719219675227 & 0.0377  & &           &       S & ElisabethB\\
SDSS J133227.20+112910.4 & 588017570316615795 & 0.0778 & Sy  2   &                 &    S & veggy2\\
SDSS J133416.49+311709.1 & 587739609171230755 & 0.0570 & SB & Was 75    &     T \\
SDSS J133718.72+242303.3 & 587742190986657795 & 0.1076 & NLSy1  &    &  S        & c\_cld \\
SDSS J133815.86+043233.3 & 587729158970736727 & 0.0228 & Sy 1.5 & NGC 5252              &S T& laihro \\
SDSS J133817.11+481636.1 & 587732483292266549 & 0.02786 & Sy 2 &  NGC 5256, Mkn 266          & S T & Gumbosea \\
SDSS J134442.16+555313.5 & 587735666377949228 & 0.0373  & Sy 2   & Mkn 273                 & S T & stellar190 \\
SDSS J134608.10+293810.4 & 587739504478060626 & 0.0776 & Sy 1 &               &     S  & lovethetropics\\
SDSS J134630.29+283646.3  & 587739707943092392 & 0.0518 &  Sy 2  &              &   T \\
SDSS J135255.67+252859.6 & 587739810484650051 & 0.06387 & Sy 1 & KUG1350+257  &  T \\
SDSS J135602.62+182217.7 & 587742550676275314 & 0.0506 & Sy 2   & Mkn 463             &   T \\
SDSS J135635.73+232135.9 & 587739845379883044 & 0.0668 & LINER &     &  T \\
SDSS J135712.06-070433.0 & 587746236298231865  & --- &      &           &      S & stellar190\\
SDSS J140037.11+622132.7 & 587728918446407773 & 0.0752  & Sy 1 &                   &    S & mitch\\
SDSS J141051.82+410412.5 & 588017604156457121 & 0.0812 & LINER &              &   S  & Song \\
SDSS J140506.26+024618.2 & 587726033335943373 & 0.0766 & Sy 2 &                   &        S  & mitch \\
SDSS J141405.01+263336.8 & 587739720298201117 & 0.0357 & Sy 1  &                 &     S & davidamesjones\\
SDSS J142522.28+141126.5 & 587742609727684701 & 0.0601  & Sy 2  &                          &     T \\
SDSS J142925.07+451831.8 & 587735490282848380  & 0.0748 & Sy 1.5  &    purple haze      &    S T & Aroel\\
SDSS J143029.88+133912.0 & 587736809916399664 & 0.0852  & Sy 2  &  Teacup                         &  S T &  Half65\\
SDSS J143239.83+361808.0  & 587736583892238376 & 0.0132 & Sy 2/SB & NGC 5675     &    T \\
SDSS J144038.10+533015.8 & 587733427086426161 & 0.0376 & Sy 2&  Mkn 477                &     T \\
\hline
\end{tabular}
\end{minipage}
\end{table*}

\setcounter{table}{2}

\begin{table*}
\contcaption{}
 \centering
 \begin{minipage}{140mm}
  \caption{Candidate AGN with extended emission-line clouds}
  \begin{tabular}{@{}llccccl@{}}
  \hline
Coordinate name       &       SDSS ObjID      &    $z$ &  Nucleus &  Name/note      & Search& Posted by \\
 \hline
SDSS J144240.79+262332.5 & 587739457225097282 & 0.1071 & Sy 1   &              &    S& spiralmania \\
SDSS J144331.19+191121.0 & 587742062161428638 & 0.0598  & Sy 2  &                 &  S & Bruno\\
SDSS J145724.63+105937.3 & 587736807771930760 & 0.1227 &  Sy 1   &              &    S & davidjamesjones\\
SDSS J150408.46+143123.3 & 587742575922708553 & 0.1181 & Sy 1 & Mkn 840      &  S & mitch\\
SDSS J150756.88+032037.3 & 587726100952449048 & 0.1369 &  Sy 1   &               &   S & mitch\\
SDSS J151004.01+074037.1 & 588017991773520114 & 0.0458 & Sy 2     &                       &    S T  & whitefluffydogs\\
SDSS J151141.26+051809.2 & 587736546312323142 & 0.0845 & Sy  1    &                 &    S& Half65 \\
SDSS J151915.98+104847.8 & 587736813131989104 & 0.0988 & Sy 1   &                   & S T & spiralmania\\
SDSS J152412.58+083241.2 & 588017703489372418 & 0.0371 & Sy 2    &                       &  S T & Alice  \\
SDSS J152549.54+052248.7 & 587730022796755031 & 0.048 & Sy 2  &                        &  T \\
SDSS J152907.45+561606.6 & 587742882456731737 & 0.0998 & Sy 1  &                 &  S& spiralmania \\
SDSS J153355.15+585756.4 & 587725818571063416  & --- &          &           &    S & lovethetropics\\
SDSS J153432.52+151133.2 & 587742013841145937 & 0.0066 & Sy 2  & NGC 5953  &   S    & Half65 \\    
SDSS J153508.93+221452.8  & 587739814240190581 & 0.0858 & trans & purple haze  &  T \\
SDSS J153703.36+135944.1 & 587742590401904799 & 0.0737 &LINER &                       &  S & RandyC\\
SDSS J153854.16+170134.2 & 587739845390761994 & 0.02974& Sy 2 &   NGC 5972                          &  S  T& NeilGibson  \\
SDSS J155007.62+272814.5 & 587736941990969374 & 0.1468 & Sy 1 & purple haze      &    S & ElisabethB\\
SDSS J160536.79+174807.5 & 587739720846934175 & 0.0339 & Sy 2  & IC 1182                       & S  & stellar190\\
SDSS J162538.08+162718.1 & 587739814246023211 & 0.0343 & LINER & Akn 502    &  T \\
SDSS J162804.06+514631.4 & 587736980102643827 & 0.0547 & Sy 1.9 & Mkn 1498            &    S & Budgieye\\
SDSS J162930.01+420703.2 & 587729653421441105 & 0.0717 & Sy 1 & purple haze      &    S & Tsering\\
SDSS J162952.88+242638.4 & 587736898503639075 & 0.0368 & Sy 1  & Mkn 883                &  S T & Rick Nowell\\
SDSS J164800.81+295657.4 & 587733399186898947 & 0.1059  & Sy 1   &               &   S & mitch\\
SDSS J172335.75+342133.4 & 587739849686843709  & --- &          &                    &            S& Mukund Vedapudi \\
SDSS J172747.17+265121.4 & 587729409160183880 & 0.0291 & &  VV 389  & S     &  elizabeth  \\
SDSS J172935.81+542939.9 & 587725505559855518 & 0.0820 & Sy 2 &                 &     S & Bruno\\
SDSS J181611.61+423937.3 & 758879745074397535 & 0.04120 & Sy 2 &   UGC 11185       &  S & stellar190\\
SDSS J210918.38-060754.7 & 587726879412256901 & 0.0286 & Sy  2   &                               & S & mitch\\
SDSS J214150.10+002209.4 & 587731186725683280 & 0.1068 & LINER &     &  S & echo-lily-mai\\    
SDSS J220141.64+115124.3 & 587727221400862869 & 0.0296 & Sy 2   &                        & S T & stellar190\\ 
SDSS J233254.46+151305.4 & 587730774959652922 & 0.2148 & Sy 1     &                          &  S & ElisabethB\\
SDSS J234413.61+004813.9 & 587731187275923676 & 0.0497 & LINER &                        &  S & davidjamesjones\\
\hline
\end{tabular}
\end{minipage}
\end{table*}

\begin{table*}
 \centering
  \caption{Spectroscopic data}
  \begin{tabular}{@{}lccrrr@{}}
 Telescope & UT Dates & Range, \AA\  & Resolution, \AA\  & Slit scale "/pixel  & Galaxies observed \\
\hline
KPNO 2.1m & 2010 June 15-21 & 3630-5700 &   3.2  & 0.78  & 33 \\
Lick 3m   & 2010 July 12-15 & 5450-8260 &   4.5  & 0.78  & 11 \\
          &                 & 3495-5605 &   1.5  & 0.43  & 11 \\
Lick 3m   & 2010 Dec 1 - 3  & 4630-7410 &   4.3  & 0.78  & 13 \\
          &                 & 3280-4595 &   2.7  & 0.43  & 13 \\
Lick 3m   & 2009 Dec 17     & 5250-9940 &  13.5  & 0.78  &  2 \\
          &                 & 3650-5710 &   4.8  & 0.43  &  2 \\

\hline
\end{tabular}
\label{tbl-observations}
\end{table*}

\begin{table*}
 \centering
  \caption{Results of long-slit spectroscopy}
  \begin{tabular}{@{}lccccrcc@{}}
 Coordinate name      &      SDSS ObjID   &      $z$  &   Nucleus  &  Name       &       PA$^\circ$	& Source 	& Region Type  \\
 \hline
 Confirmed AGN-ionized: \\
SDSS J074241.70+651037.8 & 758878270293868614 & 0.0371 & Sy 2 &  Mkn 78         &     90       &     Lick    &  AGN  \\  
SDSS J095559.88+395446.9 & 588016528244670522 & 0.0483$^*$ & Sy  2$^*$ &             &       148    &        Lick &   AGN \\
SDSS J100507.88+283038.5 & 587741392112451744 & 0.0517 & Sy 2 &                  &    62        &    Lick  & AGN      \\
SDSS J111349.74+093510.7 & 587734892748144649 & 0.0292 & Sy 1.5 & IC 2637     &        47    &        GCam      &  AGN \\
SDSS J113629.36+213551.7 & 587742013279502427 & 0.0297  & Sy 1 & Mkn 739   &           168       &     GCam     &   AGN \\
SDSS J121819.30+291513.0 & 587739719750058064 & 0.0477  & Sy 2 &  UGC 7342     &      133    &        GCam   &     AGN \\
SDSS J122546.72+123942.7 & 588017566564155399 & 0.0086 & Sy 2 & NGC 4388     &         26    &        GCam   &     AGN \\
SDSS J133815.86+043233.3 & 587729158970736727 & 0.0228 & Sy 1.5 & NGC 5252         &  175       &     GCam & AGN  \\
SDSS J133817.11+481636.1 &  587732483292266549 & 0.0279 & Sy 2 &  Mkn 266 & 176         &   GCam    Lick & AGN \\
SDSS J134442.16+555313.5 & 587735666377949228 & 0.0373  & Sy 2 &  Mkn 273       &      57    &        GCam  &    AGN \\
SDSS J135602.62+182217.8 & 587742550676275314 & 0.0504 &  Sy 2 &  Mkn 463    &  8   & GCam & AGN \\
SDSS J143029.88+133912.0 & 587736809916399664 & 0.0852 &  Sy 2 &  Teacup   &   37      &      GCam Lick  &   AGN \\
SDSS J151004.01+074037.1 & 588017991773520114 & 0.0458  & Sy 2 &         &            175    &        GCam  &  AGN \\
SDSS J152412.58+083241.2 & 588017703489372418 & 0.0371 & Sy 2 & CGCG 077-117    &     150    &        GCam & AGN \\ 
SDSS J153854.16+170134.2 & 587739845390761994 & 0.0297 & Sy 2 &  NGC 5972   &        167    &        GCam   Lick &  AGN \\ 
SDSS J162804.06+514631.4 & 587736980102643827 & 0.0547 & Sy 1.9 & Mkn 1498     &       150    &        GCam    Lick &  AGN \\ 
SDSS J162952.88+242638.4 & 587736898503639075 & 0.0368 &  Sy 1 &  Mkn 883       &      91      &      Lick    &  AGN \\
SDSS J181611.61+423937.3 & 758879745074397535 & 0.0412  & Sy 2 &  UGC 11185    &       90    &        GCam     Lick  &  AGN \\  
SDSS J220141.64+115124.3 & 587727221400862869 & 0.0296 &  Sy 2 &                      &  19         &   GCam     Lick &   AGN \\
                                 \\
Other cloud types: \\
SDSS J003507.44+004502.1 & 587731187281494175 & 0.1205 & LINER  &             &     35        &    Lick & H II regions \\
SDSS J005607.66+254804.7 & 587740589487030353 & 0.1530 & Sy 1 &      &      200   &        Lick  &     no  \\
SDSS J012839.87+144553.8 & 587724233179660360 & 0.0452 & Sy 2 &  CGCG436-060    &     45       &     Lick    &    no \\
SDSS J014238.47+000514.7 & 588015509280587804 & 0.1459 & Sy 1 &                  &     80    &        Lick &  AGN \\
SDSS J030639.58+000343.2 & 588015509289762862 & 0.1074 & Sy 1&                     &  40     &       Lick & AGN (small) \\
SDSS J033013.26-053235.9 & 587724242842026028  & 0.0131 & SB$^*$ &  NGC 1346        &      76      &    Lick & H II \\
SDSS J040548.78-061925.7 & 587727178476093634 & 0.0556 & Sy 1 &              &   76     &       Lick &   H II \\ 
SDSS J080452.73+212050.2 & 588016878287650850 & 0.1242 & Sy 1 &               &   150     &       Lick  &   AGN (small) \\  
SDSS J082642.63+111555.5 & 587745244691628370 & 0.0884 & trans &     &  72        &    Lick & resolved?   \\
SDSS J083525.51+104925.7 & 587744873714679862 & 0.1172 & Sy 1 &         &      150  &  Lick &  small? \\
SDSS J084344.98+354942.0 & 587732484342415393 & 0.0539 & Sy 2 &             &        141    &        Lick & AGN \\
SDSS J111100.60-005334.9 & 588848898833580220 & 0.0904 & Sy 2 &               &         101    &        GCam  &   unresolved cloud at 4" \\
SDSS J113323.97+550415.8 & 587733081347063838 & 0.0085 & Sy 1 & Mkn 177   &           136    &        GCam  & AGN in small compn \\
SDSS J114454.85+194635.3 & 588023669168537695 & 0.0274 & Sy 2  &                 & 131 & GCam &  unresolved $< 2$"  \\
SDSS J121418.25+293146.7 & 587741532253519916 & 0.0632 &  Sy 2 &  Was 49ab     &       63      &     GCam   & off-nuc AGN or cloud \\
SDSS J123046.11+103317.3 & 587732772131504164 & 0.01540   & SB   &   VPC 0764    &  30      &      GCam    &    unresolved \\
SDSS J124450.84-042604.5 & 587745544806727722  &  0.0147$^*$   &  LINER$^*$ &  IC 0812 &    62       &     GCam & $< 2$" \\
SDSS J125741.04+202347.7 & 588023670249750583 & 0.0807  & Sy 2 &  IC 3929  &    47    &         GCam & H II  to 9"\\
SDSS J134630.29+283646.3  & 587739707943092392 & 0.0518 &  Sy 2  &              &  164 &  GCam & 34" dim AGN cloud? \\
SDSS J135255.67+252859.6 & 587739810484650051 & 0.06387 & Sy 1 & KUG1350+257  &  163 & GCam &  H II \\
SDSS J142522.28+141126.5 & 587742609727684701 & 0.0601  & Sy 2  &                          &    110 & GCam & two AGN?  \\
SDSS J150408.46+143123.3 & 587742575922708553 & 0.1181 & Sy 1 &  Mkn 840     &        70      &      GCam  &  ~7" \\
SDSS J151915.98+104847.8 & 587736813131989104 & 0.0099 & Sy 1 &   & 89      &      GCam & unresolved $< 2$" \\
SDSS J153508.93+221452.8  & 587739814240190581 & 0.0858 & trans &                      &               70        & GCam & $< 2$"  \\
SDSS J153703.36+135944.1 & 587742590401904799 & 0.0737 & LINER &  &  21       &     GCam   & unresolved  \\
SDSS J160536.79+174807.5 & 587739720846934175 & 0.0339  & Sy 2 &  IC 1182   &  96       &     GCam   &   H II  \\
SDSS J210918.38-060754.7 & 587726879412256901 & 0.0286  & Sy 2 &                &      70       &      GCam     & unresolved $< 2$" \\
SDSS J214150.10+002209.4 & 587731186725683280 & 0.1068 & LINER &               &  70 &      GCam     Lick & unresolved $< 2$"     \\
SDSS J233254.46+151305.4 & 587730774959652922 & 0.2148 &  Sy 1 &                  &     138    &      Lick & unresolved $< 2$" \\
SDSS J234413.61+004813.9 & 587731187275923676 & 0.0497 & LINER &           &   71    &           Lick & H II \\
\hline
\end{tabular}
\label{tbl-confirmations}
\end{table*}

\begin{table*}
 \centering
  \caption{Emission-line ratios and selected fluxes}
  \begin{tabular}{@{}lrcccccccccc@{}}
\multispan2 Object    &   ${ {\rm{[Ne~V]} \lambda 3426} \over {{\rm [Ne~III]} \lambda 3869}} $ 
& ${ {\rm [Ne~III] \lambda 3869}  \over {\rm [O ~II] \lambda 3727}}$ &
$ {{{\rm [O~II]} \lambda 3727} \over {{\rm [O~III]} \lambda 5007}}$ & 
${{{\rm He~II} \lambda 4686 }\over{\rm H\beta}}$ &
 ${{{\rm [O~III]} \lambda 5007} \over{\rm H\beta}}$ &
  F(5007)  &
 ${{{\rm  [O I]} \lambda 6300} \over {{\rm H}\alpha}}$ &
 ${{{\rm  [N II]} \lambda 6583} \over {{\rm H}\alpha}}$ &
 ${{{\rm  [S II]}} \over {{\rm H}\alpha}}$ &
  F(H$\alpha$) \\
\hline
IC 2637 & nuc   &    0.23      &      0.28    &       0.81    &    0.11    &     3.94     &   2.7e-14 \\
       & cloud  & $ <0.2$        &     0.12       &    2.43    &  $ <0.06$  &       1.34     &   4.6e-15 \\
Mkn  78 & nuc (1)  & 0.35      &      0.32       &    0.10     &   0.29     &   13.5     &    6.8e-14    &  0.05        &     1.00       &     0.23   &      6.5e-13 \\
        & cloud    & 2.00        &    0.16    &       0.43   &     0.37   &     14.00   &     7.8e-15  &    0.094     &       0.51     &       0.22    &     2.0e-15 \\
Mkn  266 & nuc  &     0.69        &    0.17       &    0.43     &   0.19      &   4.08     &   3.4e-14    &  0.052       &     0.58       &     0.35     &    4.0e-14 \\
        & cloud   &  0.46      &      0.40     &      0.22   &     0.38   &      8.20   &     3.7e-14  &    0.029    &        0.25    &        0.23   &      1.1e-14 \\
Mkn  273 & nuc  &     0.27       &     0.11     &      0.90    &    0.12    &     3.93     &   2.4e-14 \\
        & cloud   &  0.06       &     0.26    &       0.93   &     0.27    &     9.3    &     1.0e-14 \\
Mkn  463 & nuc  &     0.26      &      0.29      &     0.30      &  0.08     &    7.7     &    3.7e-13 \\
        & cloud   &  0.41       &     0.16   &        0.51  &      0.17   &     11.3   &      1.8e-14 \\
Mkn  739 & nuc (2) &  1.16       &     0.49       &    0.15     &   0.18     &    5.88     &   1.5e-14 \\
        & cloud    & 1.31    &        0.22   &        0.45  &      0.24   &      8.30  &      6.5e-15   \\
Mkn  883 & nuc  &     0.31     &       0.12    &       0.81  &      0.08    &     2.82   &     5.3e-14  &    0.12    &         0.47   &         0.47   &      8.4e-14 \\
        & cloud    & 0.30     &       0.22     &      0.30   &     0.19   &      7.84   &     6.6e-15  &    0.08    &         0.31    &        0.49   &      4.1e-15         \\
Mkn 1498 & nuc (2) &  0.73    &        0.82     &      0.07  &      0.10   &      3.72   &     1.5e-13  &    0.07     &        1.00    &        0.22    &     4.9e-14 \\
        & cloud   &  1.08     &       0.62     &      0.18   &     0.32   &      12.3  &      1.0e-14 &     0.06     &        0.14    &        0.20   &      3.4e-15 \\
NGC 4388 & nuc  &     0.45      &      0.25    &       0.27   &     0.23    &     11.5   &     3.6e-13 \\
        & cloud   &  0.39        &    1.15    &       0.57   &   $ <0.23$   &      4.72   &     4.3e-15 \\
NGC 5252 & nuc  &     0.31     &       0.27    &       0.55  &      0.19   &      9.22   &     6.0e-14 \\
       &  cloud    & 1.31      &      0.34     &      0.31   &     0.34    &    10.1   &      1.1e-14 \\
NGC 5972 & nuc  &     2.25      &      0.31    &       0.29   &     0.40  &       2.56   &     5.1e-14  &    0.07     &        0.69   &         0.49   &      1.9e-14 \\
       &  cloud   &  0.15     &       0.20    &       0.63  &      0.26    &     8.48   &     2.6e-14  &    0.12      &       0.79    &        0.61   &      9.5e-15 \\
SDSS 0955+39 & nuc & $ <0.54$        &    0.05    &       4.60  &    $ <0.04$   &      2.90   &     6.9e-16  &    0.32     &        0.84     &       1.02   &      1.6e-15 \\
         &  cloud  & 2.64      &      0.47     &      0.28   &     0.51   &      7.00  &      3.2e-15 &     0.08    &         0.26   &         0.37   &      9.5e-16     \\
SDSS1005+28 & nuc &   0.55     &       0.29    &       0.30   &     0.24   &      7.96   &     1.1e-14  &    0.07    &         0.57   &         0.48   &      5.8e-15 \\
          & cloud  & 1.52     &       0.60     &      0.24    &    0.58    &     5.46   &     4.5e-15 & $  <0.01$     &        0.25    &        0.34   &      3.3e-15  \\
SDSS 1510+07 & nuc &  0.04       &     1.15    &       1.86 &     $ <0.01$ &         2.11   &     3.2e-15 \\
          & cloud  & 1.76      &      0.22    &       0.56   &     0.39   &      6.64    &    4.4e-15 \\
SDSS 1524+08 & nuc &  0.78       &     0.19    &       0.91   &     0.37   &      8.00   &     4.8e-15 \\ 
          & cloud  & 1.58     &       0.15     &      0.49   &     0.44    &     7.84    &    5.4e-15 \\
SDSS 2201+11&  nuc &  0.17    &        0.07    &       1.65  &      0.28    &     5.48   &     1.7e-15  &    0.17      &       1.21    &        1.02   &      7.0e-15 \\
          & cloud  & 0.53     &       0.24    &       0.21   &     0.18   &     10.1   &      7.8e-15  &    0.07    &         0.40   &         0.35   &      3.0e-15 \\
Teacup    &   nuc  & 0.31     &       0.23    &       0.30   &     0.09   &      7.95   &     1.3e-13 &     0.15     &        0.23    &        0.15    &     1.1e-14 \\
          & cloud  & 0.14      &      0.17    &       0.23   &     0.15   &      8.07  &      2.1e-14  &    0.11     &        0.34    &        0.36   &      7.6e-15 \\
UGC 7342  &   nuc &  0.75     &       0.22   &        0.58   &     0.22   &      9.14   &     2.9e-14 \\
         &  cloud  & 0.77     &       0.26    &       0.41   &     0.55   &      9.90   &     8.9e-15 \\
UGC 11185  &  nuc (1) & 0.17    &      0.19    &       0.37  &      0.16   &      8.89  &      8.5e-14 &     0.19   &          1.29   &         0.88   &      3.8e-14 \\
          & cloud  & 0.28      &      0.28    &       0.25   &     0.26   &      8.11   &     1.4e-14  &    0.07     &        1.10    &        0.72   &      5.1e-15 \\
\hline
\end{tabular}
Notes: Line fluxes are in units of erg cm$^{-2}$ s$^{-1}$ \\
(1) blend of two velocity components\\
(2) BLR present; flux is estimated NLR only
\label{tbl-lineratios}
\end{table*}

\begin{table*}
 \centering
  \caption{Ionizing and far-infrared luminosity comparison}
  \begin{tabular}{@{}lccccccccccc@{}}
 SDSS ID   & Other name &  IRAS $60 \mu$  & $100 \mu$ & Akari  $90 \mu$ & $140 \mu$ &       L(FIR) & $r$" & F(H$\beta$) & $\alpha^\circ$ & $L_{ion}$ &    Ratio \\
\hline
074241.70+651037.8  &   Mkn 78      &  1.11  & 1.13 & $0.9 \pm 0.2$ & $1.4 \pm 0.1$ &  2.4e44         &   10   &  4.0e-16  &   11 &  $>5.7$e43 &    0.25 \\
095559.88+395446.9  &                     &  -- & -- & $1.2 \pm 0.2$ & $2.5 \pm 0.3$  &   5.0e44  &  7   &  2.5e-16    &  16 &   $>3.0$e43  &   0.06 \\ 
100507.88+283038.5  &                       & -- & -- & -- & -- & $ <2.1$e44       & 8   &  4.2e-16   &    14 &  $>1.0$e44   &    0.5 \\
111349.74+093510.7  & IC 2637        & 1.75 & 3.39 & $2.6 \pm 0.1$ & $4.0 \pm 0.7$  & 3.0e44  &   10  &   3.0e-16   &  11 &   $>2.7$e43  &      0.10\\
113629.36+213551.7 & Mkn 739       &  1.3 &  2.4   &    $1.7 \pm 0.07$ & $3.5 \pm 0.6$ & 2.3e44 & 18 & 2.1e-16 & 6.3 & $>6.1$e43 & 0.26 \\
121819.30+291513.0  & UGC 7342   &  $<0.2$ & $<0.6$ & -- & -- &   $<1.1e$44  &   35  &   1.8e-16    &  43.2 &   $>4.4$e44  &    3.6 \\
122546.72+123942.7  & NGC 4388   &   10.0 & 17.1 & $10.4 \pm 0.6$ & $15 \pm 3$ & 1.4e44   &  15   &  3.8e-16   &   7.6 & $>6.3$e42   &    0.02\\
133815.86+043233.3  & NGC 5252   &    $<0.2$ & $<0.6$ & $ 0.4 \pm 0.1$ & -- &  4.0e43  &   32   &  3.8e-16  &   3.5  & $>2.0$e44   &   5 \\  
133817.11+481636.1 & Mkn 266        &  7.3 & 10.3 & $7.1 \pm 0.3$  & $9.0 \pm 1.2$ & 1.0e45   &  27  &   6.8e-16  &    4.2  & $>3.9$e44   &    0.4\\
134442.16+555313.5 & Mkn 273        &  22.5  & 22.5    & $20.2 \pm 0.6$ & $14.3 \pm 1.1$ &  5.0e45  &   12  &   9.0e-16  &   10  & $>1.6$e44   &    0.03\\
135602.62+182217.8  & Mkn 463       & 2.33 & 1.94 & $1.58 \pm 0.03$ & $2.2 \pm 0.5$ & 9.0e44    & 16   &  3.8e-16    &  7  &  $>2.5$e44   &    0.3\\
143029.88+133912.0 & Teacup          & 0.26 & $<0.6$ & -- & -- & $<2.3$e43  & 5   &  1.66e-15  &  22   & $>3.2$e44  &   14 \\
151004.01+074037.1 &                       &   $<0.2$ & $<0.6$ & -- & -- &  $<4$e44    &   14  &   8.2e-16   &   8   & $>1.3$e45  &     3.4 \\
152412.58+083241.2 &                       &    0.71     &   0.73   & 0.63 & $<0.96$ &   1.6e44  & 5.6e-16  &  12 &    10  &  1.0e44 & 0.6 \\
153854.16+170134.2 & NGC 5972    &  0.24 & $<0.76$  & -- & -- & $<5.5$e43  &   35  &   7.2e-17  &    3   & $>7.8$e43  &      1.8\\
162804.06+514631.4 & Mkn 1498     &  0.34 & $<0.61$  & -- & -- & $<5.$e43    &  10   &  9.8e-17  &   11    & $>3.1$e43  &   0.6 \\
162952.88+242638.4 & Mkn 883       & 1.01 & 1.13 & $0.84 \pm 0.04$ & -- & 2.2e44   &  10   &  4.0e-16    & 11  &  $>5.9$e43  &     0.26\\
181611.61+423937.3 & UGC 11185  & $<0.4$ & $<0.8$ & -- & --  & $<1.8$e44  &   10  &   1.7e-15  &   11  & $>3.1$e44  &    1.7 \\
220141.64+115124.3 &                       & 0.28 & $<0.99$ & -- & -- & $<6.1$e43  &   23  &   4.6e-16  &    5.0 & $>2.0$e44  &  3.4 \\
\hline
\end{tabular}
Notes: FIR fluxes are in Jy \\
Mean IRAS detection limits are used when no specific value is available\\
Luminosities are in erg s$^{-1}$; H$\beta$ fluxes are in erg cm$^2$ s$^{-1}$\\
Values of ionizing/FIR luminosity ratio are all lower limits
\label{tbl-energy}
\end{table*}

\begin{table*}
 \centering
  \caption{[S II] density measures and limits}
  \begin{tabular}{@{}lccrccccc@{}}
Object & Distance: arcsec &  kpc  & [S II] ratio & $n_e$ (cm$^{-3}$) & [O II]/[O III] & log $U$ & $L_{ion}$ (erg/s)\\
\hline
Teacup &   5-14 & 8-22  &     $ 1.25 \pm 0.04$ &    130-240 & 0.15 & -2.16 & $<1.9e46$\\
Mkn 883  & 12-16 & 9-12    &  $ 1.36 \pm  0.03$ &   12- 100 & 2.62 & -3.34 & $<4.8e44$ \\
SDSS 2201& 8-16 & 5-10 &  $ 1.48 \pm  0.06$ &  $  <10$ & 0.25 & -2.42 & $<2.2e44$ \\
NGC 5972  & 26-30 &15-18 & $  1.29  \pm  0.12$ &   3-300 & 0.11 & -1.99 & $<9.8e46$ \\
Mkn 266 & 20-29 & 11-16 &  $ 1.36 \pm  0.04$ &   12-100 & 0.22 & -2.36 & $<9.3e45$ \\
\hline
\end{tabular}
\label{tbl-s2density}
\end{table*}

\begin{table*}
 \centering
  \caption{Morphologies of AGN hosts with extended clouds}
  \begin{tabular}{@{}lcccclccc@{}}
 SDSS designation                      & $z$     & Sy type & Name    & $r_{max}$, kpc & Morphology &   cone angle$^\circ$ &  disc/cloud angle$^\circ$ &  Sides \\
\hline
SDSS J074241.70+651037.8  & 0.0371 & 2  &  Mkn 78        & 16   &   E                               &         55          &        &              2 \\     
SDSS J095559.88+395446.9  & 0.0483  & 2  &                     & 10     &  Interacting S           &           88        &             &                         1 \\  
SDSS J100507.88+283038.5  & 0.0517  & 2   &                      & 13  & Sb, disturbed companion      &          92 &         &                             1 \\ 
SDSS J111349.74+093510.7  & 0.0292  & 1.5 & IC 2637     & 11  &  Merger remnant           &          60 &         &                    1 \\  
SDSS J113629.36+213551.7  & 0.0297  & 1 &  Mkn 739     & 17   & Ongoing merger  &     28 &       & 1     \\ 
SDSS J121819.30+291513.0  & 0.0477  & 2   & UGC 7342  & 38   & Ongoing merger;  tails       &           86 &        &       2 \\    
SDSS J122546.72+123942.7  & 0.0086  & 2  & NGC 4388   & 13  &  Edge-on Sc  & 80 &  53    &       1 \\     
SDSS J133815.86+043233.3  & 0.0228  & 1.5 & NGC 5252 & 21  & Edge-on S0, tilted H I ring       &      59 &     31     &      2 \\
SDSS J133817.11+481636.1  & 0.0279 & 2   & Mkn 266    & 21     &  Ongoing merger             &      112 &             &                       2 \\ 
SDSS J134442.16+555313.5  & 0.0373 &  2  & Mkn 273      & 19   &  Ongoing merger            &           75 &        &                           2 \\
SDSS J135602.62+182217.8 & 0.0504 &  2 &  Mkn 463E  & 16   &  Ongoing merger     &         55 &                            &                2  \\ 
SDSS J143029.88+133912.0  & 0.0852  & 2  & Teacup       & 18    &  Stellar tail and arc          &          80 &         &                          1 \\ 
SDSS J151004.01+074037.1  & 0.0904  & 2   &                    & 10    &  Symmetric disc; S0 or Sa      &        85 &           &                   2 \\ 
SDSS J152412.58+083241.2 &  0.0371 & 2 & CGCG 077-117 & 19 & Merger remnant         &        56   &             &                        1 \\  
SDSS J153854.16+170134.2  & 0.0297 & 2  & NGC 5972   & 33   &  Warped disc and tails         &    35     &   18       &                  2 \\ 
SDSS J162804.06+514631.4  & 0.0547 & 1.9 &  Mkn 1498   & 21  &  E                        &         42 &          &                                          1 \\ 
SDSS J162952.88+242638.4  & 0.0368  & 1  & Mkn 883      & 37   &  Ongoing merger         &         73 &          &                               1 \\  
SDSS J181611.61+423937.3  & 0.0412 & 2  & UGC 11185   & 11 &  Strong interaction       &         48 &           &                              1 \\ 
SDSS J220141.64+115124.3  & 0.0296  & 2    &                   & 16    &  Edge-on warped disc, tails   &  23 &    30   &       2 \\  
SDSS J094104.11+344358.4  & 0.0499  & LINER    & IC 2497 & 40 & Warped disk, H I tail &     46    &    65       &     1 \\     
\hline
\end{tabular}
\label{tbl-morphology}
\end{table*}

\label{lastpage}


\begin{thebibliography}{99}
\bibitem[\protect\citeauthoryear{Alonso et al.}{2007}]{Alonso} Alonso, M.~S., Lambas, D.~G., Tissera, P., \& Coldwell, G.\ 2007, MNRAS 375, 1017 
\bibitem[\protect\citeauthoryear{Antonucci}{1993}]{Ski93} Antonucci, R. 1993, ARA\&A 31, 473
\bibitem[\protect\citeauthoryear{Baldwin, Phillips, \& Terlevich}{1981}]{BPT} Baldwin, J.A., Phillips, M.M. \& Terlevich, R. 1981, PASP 93, 5
\bibitem[\protect\citeauthoryear{Barbosa et al.}{2009}]{Barbosa} Barbosa, F.K.B., Storchi-Bergmann, T., Cid Fernandes, R., Winge, C. \& Schmitt, H. 2009, MNRAS 396, 2
\bibitem[\protect\citeauthoryear{Bennert}{2005}]{Bennert2005} Bennert, N. 2005, Ph.D. dissertation, Ruhr-Universit\"at Bochum
\bibitem[\protect\citeauthoryear{Bennert et al.}{2006a}]{Bennert2006a} Bennert, N., Jungwiert, B., Komossa, S., Haas, M. \& Chini, R. 2006a, A\&A 446, 919
\bibitem[\protect\citeauthoryear{Bennert et al.}{2006b}]{Bennert2006b} Bennert, N., Jungwiert, B., Komossa, S., Haas, M. \& Chini, R. 2006b, A\&A 456, 953
\bibitem[\protect\citeauthoryear{Bianchi et al.}{2008}]{Bianchi} Bianchi, S., Chiaberge, M., Piconcelli, E., Guainazzi, M. \& Matt, G. 2008, MNRAS 386, 105 
\bibitem[\protect\citeauthoryear{Capetti et al.}{1994}]{Capetti1994} Capetti, A., Macchetto, F., Sparks, W. B., \& Boksenberg, A. 1994, ApJ 421, 87
\bibitem[\protect\citeauthoryear{Chatzichristou \& Vanderriest}{1995}]{Chatzichristou} Chatzichristou, E.T. \& Vanderriest, C. 1995, A\&A 298, 343
\bibitem[\protect\citeauthoryear{Condon \& Broderick}{1988}]{Condon88} Condon, J.J. \& Broderick, J.J 1988, AJ 96, 30
\bibitem[\protect\citeauthoryear{Condon et al.}{1998}]{NVSS} Condon, J.J., Cotton, W.D., Greisen, E.W., Yin, Q.F., Perley, R.A., Taylor, G.B. \& Broderick, J.J.
1998, AJ. 115, 1693
\bibitem[\protect\citeauthoryear{Cracco et al.}{2011}]{Cracco} Cracco, V. et al. 2011, MRAS (in the press), arXiv:1109.1195
\bibitem[\protect\citeauthoryear{Dadina et al.}{2010}]{Dadina} Dadina, M., Guainazzi, M., Cappi, M., Bianchi, S., Vignali, C., Malaguti, G., \& Comastri, A.\ 2010, A\&A 516, A9 
\bibitem[\protect\citeauthoryear{da Silva et al.}{2011}]{daSilva} da Silva, R.L., Prochaska, J.X., Rosario, D., Tumlinson, J., \& Tripp, T.R. 2011, ApJ 735, 54
\bibitem[\protect\citeauthoryear{Dopita \& Sutherland}{1996}]{Dopita96} Dopita, M.A. \& Sutherland, R.S. 1996, ApJS 102, 161
\bibitem[\protect\citeauthoryear{Fesen et al.}{1982}]{Fesen82} Fesen, R.~A., Blair, W.~P., \& Kirshner, R.~P.\ 1982, ApJ 262, 171 
\bibitem[\protect\citeauthoryear{Filippenko}{1982}]{AVF82} Filippenko, A.V. 1982, PASP 94, 715
\bibitem[\protect\citeauthoryear{Fischer et al.}{2011}]{Fischer} Fischer, T. C., Crenshaw, D. M., Kraemer, S. B., Schmitt, H. R., Mushotsky, R. F., \& Dunn, J. P.
2011, ApJ 727, 71
\bibitem[\protect\citeauthoryear{Fu \& Stockton}{2009a}]{FuStockton1} Fu, H. \& Stockton, A. 2009, ApJ 690, 953
\bibitem[\protect\citeauthoryear{Fu \& Stockton}{2009b}]{FuStockton2} Fu, H. \& Stockton, A. 2009, ApJ 696, 693
\bibitem[\protect\citeauthoryear{Fullmer \& Lonsdale}{1989}]{IRASgal} 
Fullmer, L. \& Lonsdale, C.J. 1989, Cataloged Galaxies and Quasars Observed in the IRAS Survey, Version 2, Jet Propulsion Laboratory (JPL D-1932)
\bibitem[\protect\citeauthoryear{Josza et al.}{2009}]{Josza2009} Josza, G.I.G. et al. 2009, A\&A 500, l33
\bibitem[\protect\citeauthoryear{Kauffmann et al.}{2003}]{Kauffmann2003} Kauffmann, G. et al.\ 2003, MNRAS,  346, 1055
\bibitem[\protect\citeauthoryear{Kawada et al.}{2007}]{AkariFIS} Kawada, M. et al. 2007, PASJ 59, S389--S400
\bibitem[\protect\citeauthoryear{Keel}{1983}]{Keel83} Keel, W.C. 1983, ApJ  269, 466
\bibitem[\protect\citeauthoryear{Keel}{1996}]{Keel1996} Keel, W.C. 1996, ApJS  106, 27
\bibitem[\protect\citeauthoryear{Keel et al.}{2011}]{Keel2011} Keel, W.C., et al. \ 2011,  in preparation 
\bibitem[\protect\citeauthoryear{Kewley et al.}{2001}]{Kewley2001} Kewley, L.~J., 
Dopita, M. A., Sutherland, R. S., Heisler, C. A., \& Trevena, J. 2001, ApJ, 556, 121
\bibitem[\protect\citeauthoryear{Komossa \& Schulz}{1997}]{KomossaSchulz}  Komossa S. \& Schulz, H. 1997 A\&A 323, 31
\bibitem[\protect\citeauthoryear{Koss et al.}{2011}]{Koss} Koss, M., Mushotzky, R., Treister, E., Veilleux, S., Vasudevan, R., Miller, N., Sanders, D.B.,
Schawinski, K., \& Trippe, M 2011, ApJL 735, L42
\bibitem[\protect\citeauthoryear{Kuo et al.}{2008}]{Kuo} Kuo, C.-Y., Li, J., Tang, Y.-W., \& Ho, P.T.P. 2008, ApJ 679, 1047
\bibitem[\protect\citeauthoryear{Li et al.}{2008}]{Li}  Li, C., Kauffmann, G., 
Heckman, T.~M., White, S.~D.~M., \& Jing, Y.~P.\ 2008, MNRAS 385, 1915 
\bibitem[\protect\citeauthoryear{Lintott et al.}{2008}]{Lintott2008} Lintott, C.~J., et al.\ 2008, MNRAS, 389, 1179 
\bibitem[\protect\citeauthoryear{Lintott et al.}{2009}]{Lintott2009} Lintott, C.~J., et al.\  2009, MNRAS, 399, 129
\bibitem[\protect\citeauthoryear{Lupton et al.}{2004}]{lupton} Lupton, R., Blanton, M.~R., Fekete, G., Hogg, D.~W., O'Mullane, W., Szalay, A., 
\& Wherry, N.\ 2004, PASP, 116, 133 
\bibitem[\protect\citeauthoryear{Maia et al.}{2003}]{Maia} Maia, M.~A.~G., Machado, R.~S., \& Willmer, C.~N.~A.\ 2003, AJ 126, 1750 
\bibitem[\protect\citeauthoryear{Mazzarella et al.}{1991}]{Mazzarella} Mazzarella, J.M., Soifer, B.T., Graham, J.R., Neugebauer, G., Matthews, K. \& Gaume, R.A. 1991, AJ 102, 1241
\bibitem[\protect\citeauthoryear{Moran et al.}{1992}]{Moran} Moran, E.C., Halpern, J.P., Bothun, G.D., \& Becker, R.H. 1992, AJ 104, 990
\bibitem[\protect\citeauthoryear{Morse et al.}{1995}]{Morse95} Morse, J.~A., Winkler,  P.~F., \& Kirshner, R.~P.\ 1995, AJ 109, 2104 
\bibitem[\protect\citeauthoryear{Murakami et al.}{2007}]{Akari} Murakami, H., et al., 2007, PASJ 59, S369--S376
\bibitem[\protect\citeauthoryear{Osterbrock}{1977}]{DEO77} Osterbrock, D.E. 1977, ApJ 215, 733
\bibitem[\protect\citeauthoryear{Pedlar et al.}{1989}]{Pedlar1989} Pedlar, A., Meaburn, J., Axon, D. J., Unger, S. W., Whittle, D. M., Meurs, E. J. A., Guerrine, N., \& Ward, 
M. J. 1989, MNRAS 238, 863
\bibitem[\protect\citeauthoryear{Prieto \& Freudlng}{1996}]{Prieto} Prieto, M.A. \& Freudling, W, 1996, MRAS 279, 63
\bibitem[\protect\citeauthoryear{Ramos Almeida et al.}{2006}]{Almeida} Ramos Almeida, 
C., P{\'e}rez Garc{\'{\i}}a, A.~M., Acosta-Pulido, J.~A., Rodr{\'{\i}}guez 
Espinosa, J.~M., Barrena, R., \& Manchado, A.\ 2006, ApJ 645, 148 
\bibitem[\protect\citeauthoryear{Rampadarath et al.}{2010}]{Rampadarath} Rampadarath, H., et al. \ 2010, A\&A 517, 8
\bibitem[\protect\citeauthoryear{R\"ottgering et al.}{1996}]{Rottgering} R\"ottgering, H.J.A, Tang, Y., Bremer, M.A.R., de Bruyn, A.G., Miley, G.K. Rengelink, R.B. \& Bremer, M.N. 1996, MNRAS 282, 1033
\bibitem[\protect\citeauthoryear{Schawinski et al.}{2010a}]{Schawinski2010a} Schawinski, K., et al. \ 2010a, ApJL 724, L30 
\bibitem[\protect\citeauthoryear{Schawinski et al.}{2010b}]{Schawinski2010b} Schawinski, K., et al. \ 2010b, ApJ, 711, 284 
\bibitem[\protect\citeauthoryear{Schmitt et al.}{1997}]{Schmitt1997} Schmitt, H.R., Kinney, A.L., Storchi-Bergmann, T. \&  Antonucci, R. 1997, ApJ 477, 623
\bibitem[\protect\citeauthoryear{Shaw \& Dufour}{2007}]{ShawDufour}  
Shaw, R.A. \& Dufour, R. J. 1995, PASP 107, 896
\bibitem[\protect\citeauthoryear{Spergel et al.}{2007}]{Spergel}  Spergel, D.~N., et al. 2007, ApJS 170, 377 
\bibitem[\protect\citeauthoryear{Stockton, Fu, \& Canalizo}{2006}]{Stockton2006} Stockon, A., Fu, H. \& Canalizo, G. 2006, NewAR 50, 694
\bibitem[\protect\citeauthoryear{Tadhunter \& Tsvetanov}{1989}]{Tadhunter1989} Tadhunter, C. \& Tsvetanov, Z. 1989, Nature 341, 422
\bibitem[\protect\citeauthoryear{Tody}{1986}]{Tody} Tody, D. 1986, SPIE 627, 733
\bibitem[\protect\citeauthoryear{Unger et al.}{1987}]{Unger1987} Unger, S.W., Pedlar, A., Axon, D.J., Whittle, M., Meurs, E.J.A., \& Ward, M.J. 1987, MNRAS 228, 671
\bibitem[\protect\citeauthoryear{Uomoto et al.}{1993}]{Uomoto} Uomoto, A., Caganoff, S., Ford, H.C., Rosenblatt, E.I., Antonucci, R.R.J., Evans, I.N., \& Cohen R.D. 1993, AJ 105, 1308
\bibitem[\protect\citeauthoryear{Veilleux \& Osterbrock}{1987}]{VO87} Veilleux, S. \& Osterbrock, D.E. 1987, ApJS 63, 295
\bibitem[\protect\citeauthoryear{Veron-Cetty \& Veron}{2010}]{VCV13} Veron-Cetty, M.P. \& Veron, P. A\&A 518, 10
\bibitem[\protect\citeauthoryear{Wallerstein \& Balick}{1990}]{Wallerstein} Wallerstein, G. \& Balick, B. 1990, MNRAS 245, 701
\bibitem[\protect\citeauthoryear{Whittle et al.}{2005}]{Whittle2005} Whittle, D.M., Rosario, D., Silverman, J.D., Nelson, C.H.  \& Wilson, A.S. 2005, AJ 129, 104
\bibitem[\protect\citeauthoryear{Whittle \& Wilson}{2004}]{WW2004} Whittle, D.M. \& Wilson, A.S. 2004, AJ 127, 606
\bibitem[\protect\citeauthoryear{Wilson}{1996}]{Wilson1996} Wilson, A.S. 1996, VA 40, 63
\bibitem[\protect\citeauthoryear{Wilson \& Tsvetanov}{1994}]{Wilson1994} Wilson, A.S. \& Tsvetanov, Z. 1994, AJ 107, 1227
\bibitem[\protect\citeauthoryear{Wu et al.}{2011}]{Wu2011} Wu, Y.-Z., Zhang, E.-P.,  Liang, Y.-C., Zhang, C.-M., \& Zhao, Y.-H.\ 2011, ApJ 730, 121
\bibitem[\protect\citeauthoryear{Yamamura et al.}{2010}]{AkariBSC} Yamamura, I., Makiuti, S., Ikeda, N., Fukuda, Y., Oyabu, S., Koga, T., \& White, G. J., 2010, http://www.ir.isas.jaxa.jp/AKARI/Observation/PSC/Public/RN/AKARI-FIS\_BSC\_V1\_RN.pdf
\bibitem[\protect\citeauthoryear{Yoshida et al.}{2002}]{Yoshida} Yoshida, M., et al. 2002, ApJ 567, 118 
\bibitem[\protect\citeauthoryear{Yoshida et al.}{2004}]{Yoshida2004} Yoshida, M. et al. 2004, AJ 127, 90
\end{thebibliography}
\end{document}